\documentclass[twocolumn,showpacs,preprintnumbers,amsmath,amssymb]{revtex4}

\usepackage{graphicx}
\usepackage{dcolumn}
\usepackage{bm}
\usepackage{amssymb}
\usepackage{enumerate}

\usepackage{amsmath}
\usepackage{epsf}
\usepackage{epsfig}
\usepackage{amssymb}
\usepackage{multirow}

\newcommand{\be}{\begin{equation}}
\newcommand{\ee}{\end{equation}}
\newcommand{\bear}{\begin{eqnarray}}
\newcommand{\ear}{\end{eqnarray}}

\newcommand{\ba}{\begin{array}}
\newcommand{\ea}{\end{array}}

\makeatletter

\def\lsim{\compoundrel<\over\sim}
\def\compoundrel#1\over#2{\mathpalette\compoundreL{{#1}\over{#2}}}
\def\compoundreL#1#2{\compoundREL#1#2}
\def\compoundREL#1#2\over#3{\mathrel
         {\vcenter{\hbox{$\m@th\buildrel{#1#2}\over{#1#3}$}}}}
\makeatother

\newcommand\ainv{A'\to invisible}

\newcommand\g{\gamma}
\newcommand\ma{M_{A'}}

\newcommand\na{{n}_{A'}}

\def\address{\@ifstar{\address@star}%
  {\@ifnextchar[{\address@optarg}{\address@noptarg}}}

\begin{document}

\author{S.N.~Gninenko$^{1}$}
\author{N.V.~Krasnikov$^{1,2}$}
\author{M.M.~Kirsanov$^{1}$}
\author{D.V.~Kirpichnikov$^{1}$}

\affiliation{$^{1}$ Institute for Nuclear Research of the Russian Academy of Sciences, 117312 Moscow, Russia \\
$^{2}$ Joint Institute for Nuclear Research, 141980 Dubna, Russia}



\title{Missing energy signature from invisible decays of dark photons at the CERN SPS \\
}

\date{\today}

\begin{abstract}
The dark photon ($A'$) production through the mixing with 
the bremsstrahlung photon from the  electron scattering off nuclei can be accompanied by the dominant invisible $A'$ decay  into dark-sector particles.  
In this work we discuss the missing energy signature of this process in the experiment NA64  aiming at the search for $\ainv$ decays with a high-energy electron beam at the CERN SPS.
We show the distinctive  distributions of variables that can be used to distinguish the $\ainv$ signal from background.
The results of the detailed simulation of the detector response for the events with and without $A'$ emission  are presented.
The efficiency of the signal event selection is estimated. It is used to evaluate the sensitivity of the experiment and show that it allows
to  probe the still unexplored area of the mixing strength  $10^{-6}\lesssim \epsilon \lesssim 10^{-2}$ and masses up to $M_{A'} \lesssim 1$ GeV.
The results obtained are compared with the results from other calculations. In the case of 
the signal observation, a possibility of extraction of the  parameters $M_{A'}$ and 
$\epsilon$ by using the shape of the missing energy spectrum is discussed.

 \end{abstract}
\maketitle

\section{Introduction and motivation}
The origin of dark matter is a great puzzle in the cosmology and particle physics.
In recent years, various phenomenological models assumed the existence  of  a light vector boson, the "dark photon" $A'$, with a mass $m_{A'}\lesssim 1$ GeV resulting from  a spontaneously broken new gauge symmetry $U(1)_D$. The $A'$ couples to the standard model (SM) particles  only through the kinetic mixing  of dark charge with hypercharge, parametrized by the 
mixing strength  $\epsilon \ll 1$ \cite{Okun:1982xi, Galison:1983pa,  Holdom:1985ag}. The  $A'$  kinetically mixes with the photon and couples primarily to the electromagnetic current with a strength $\epsilon e$, where $e$ is the electromagnetic coupling. The phenomenology  of  $A'$ , motivated  by potential astrophysical signals of dark matter \cite{ArkaniHamed:2008qn}, as well as the 3.6 $\sigma$ discrepancy between the the SM prediction and measurements of the muon anomalous magnetic dipole moment g-2 \cite{Bennett:2006fi} has been studied  in  
many  theoretical and experimental works  \cite{ArkaniHamed:2008qn,Bjorken:2009mm,Essig:2013lka,fayet,gk,Pospelov:2008zw}.

If the $A'$ is the lightest particle  in the dark sector, it will decay dominantly into ordinary particles, e.g. $ e, \mu$.    However, if there are lighter  
dark sector  states, $A'$ would decay predominantly  into such particles
  resulting in the $\ainv$ decay.  This will occur assuming  that  $e_D \gg \epsilon e$, where $e_D$ is the coupling constant of the $U(1)_D$ gauge interaction with light dark matter particles. Such $A'$,  which is 
nearly "invisible",  provides new possibilities to explain various anomalies including the  muon g-2 problem and  is a subject of  different experimental constraints \cite{Izaguirre:2014bca,Diamond:2013oda,Davoudiasl:2014kua,Batell:2014mga} and new experimental searches. Interestingly, that the muon (g-2)
anomaly \cite{gk,fayet,Pospelov:2008zw}  can be explained by the existence of a sub-GeV $A'$  with the couplings  $\epsilon \simeq  10^{-3}$. Such couplings naturally arise from the loop effects  of particles that are charged under both the standard model (SM)  and dark hypercharge $U(1)$ interactions \cite{Holdom:1985ag}.

One possible way  to search  for the invisible $A'$  is based on production and detection sub-GeV dark matter in accelerator experiments. The $A'$s 
produced in a high intensity beam dump  experiment, decay in flight and 
produce other dark matter particles which can be detected through the  scattering of electrons in 
the detector target \cite{Izaguirre:2014bca,Diamond:2013oda,Batell:2009di,deNiverville:2011it,Dharmapalan:2012xp}. The signal event rate depends on the $A'$ couplings to the dark and visible sectors, $e_D$ and $\epsilon e$ respectively and  scaled
 by $\epsilon^2 e_D^2/e^2$. Another approach considered in this work and proposed in Refs.\cite{Gninenko:2013rka,Andreas:2013lya}, 
 is based on the detection of the large missing energy, carried away by the energetic $A'$
 produced in the interactions of high-energy electrons in the active beam dump target, see also \cite{Izaguirre:2014bca}.   
The advantage of the second type of experiments is that their sensitivity is roughly proportional to the mixing squared, $\epsilon^2$,  associated with the $A'$ production  in the primary reaction and its subsequent prompt invisible decay, while in the former case it is proportional to $\epsilon^4$, one $\epsilon^2$ coming from the $A'$ production in the beam dump
and another $\epsilon^2$ from the cross section of the dark matter particle interactions in the active detector.

\begin{figure}[tbh!]
\begin{center}
\includegraphics[width=0.5\textwidth]{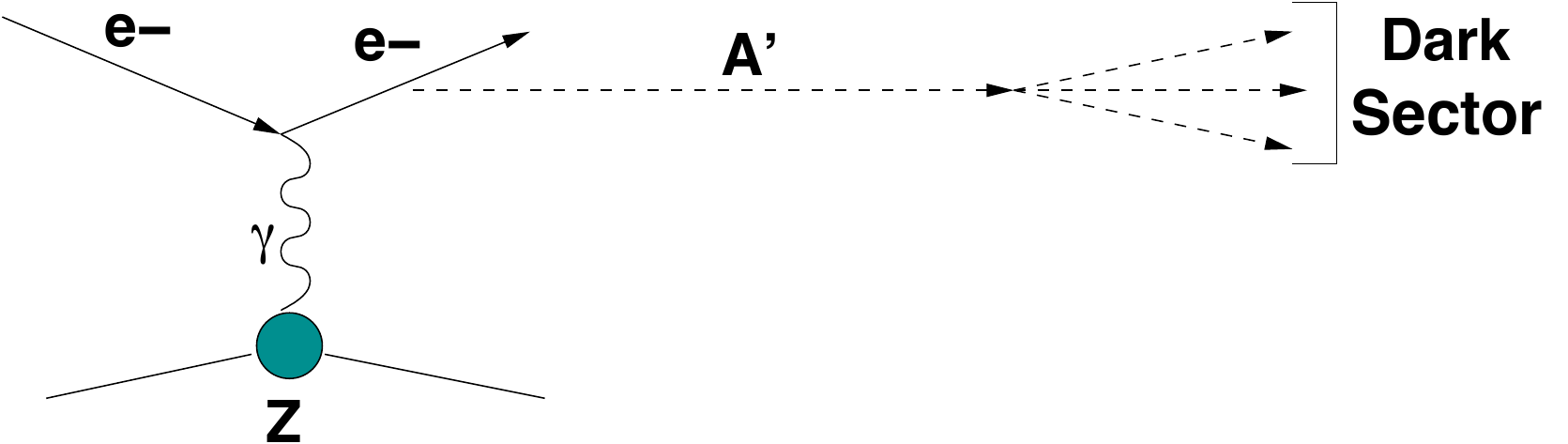}
\caption {Diagram contributing to the $A'$ production in the  reaction  
 $e^- Z \rightarrow e^-  Z A',  A' \rightarrow dark~sector$. The  produced $A'$  decays invisibly into dark sector particles.
 \label{diagr}}
\end{center}
\end{figure}


In this work we discuss the fixed-target experiment NA64 at the CERN SPS \cite{Gninenko:2013rka,Andreas:2013lya} aiming
at the search for $\ainv$ decays with a  100 GeV electron beam.
Different background sources that could mimic the signal in this experiment were studied in detail in  Refs.\cite{Gninenko:2013rka, Andreas:2013lya}, see
also \cite{Izaguirre:2014bca}. It has been shown that for the mixing $10^{-6}\lesssim \epsilon \lesssim 10^{-3}$
and masses $M_{A'} \lesssim 1$ GeV the proposed search  is expected to be  background free at the level $\lesssim 10^{-12}$ per incident electron.
Here,  we focus mainly  on the $A'$ production rate, experimental signature  of the $\ainv$ decays, and sensitivity  of the experiment. Our goal is two-fold.
 First, in light of recent disagreements in 
the literature on the question of the $A'$ yield computations
 \cite{Izaguirre:2014bca}, we revisit here the calculations of Ref.\cite{Gninenko:2013rka,Andreas:2013lya}.  We seek to clarify the apparent 
disagreements about the numerical factors in the analytic expressions for the 
$A'$ yield computations.
Obtaining a reliable  theoretical prediction for the $A'$ yieldis essential 
for the proper interpretation of the obtained experimental results  in terms of 
the possible observation of  the $A'$ signal or obtaining a robust exclusion 
limits in the $A'$ parameter space. 

 Second, we attempt to provide an estimate of the experimental  uncertainties associated with the $A'$ signal calculation required for the sensitivity estimate. While the study of Ref.\cite{Izaguirre:2014bca} included some theoretical 
uncertainties associated with the $A'$ modeling and experimental data used as 
input for the calculation, no estimate of the errors and factors  related to 
 the concrete experimental setup configuration was given. We discuss additional experimental inputs that would be useful to improve the reliability of the calculated sensitivity of the experiment. 
 We extend the analysis of Ref.\cite{Andreas:2013lya} by simulating the full detector response and taking into account the realistic production and detection  efficiency for signal events. Finally, the feasibility  of reconstruction of 
the signal parameters such as the mass and the mixing strength of the $A'$ 
  from the observed shape of the $E_{miss}$ spectrum has been studied for the 
  values $M_{A'}=20$ and 200 MeV and $\epsilon \simeq 10^{-3}$. 

The remainder of our treatment of these issues is organized as follows. 
Section II outlines the theoretical setup for the $A'$ production in 
electron- nuclei scattering, observables
 that are analyzed and the signal simulation. The results of the detector response simulation are reported in Section III. 
Section IV is dedicated to the discussion of the missing energy signature 
of the signal events $A'$ yield. The results of the detailed  detector response simulation and some background issues are reported in Section V and Section VI, respectively. In Section VII the expected sensitivity of the search is discussed and  compared with the  existing calculations  obtained by Izaguirre et al. in \cite{Izaguirre:2014bca}.
 We conclude the article with a short summary in Section VIII.


\begin{figure}[tbh]
\begin{center}
\includegraphics[width=0.5\textwidth]{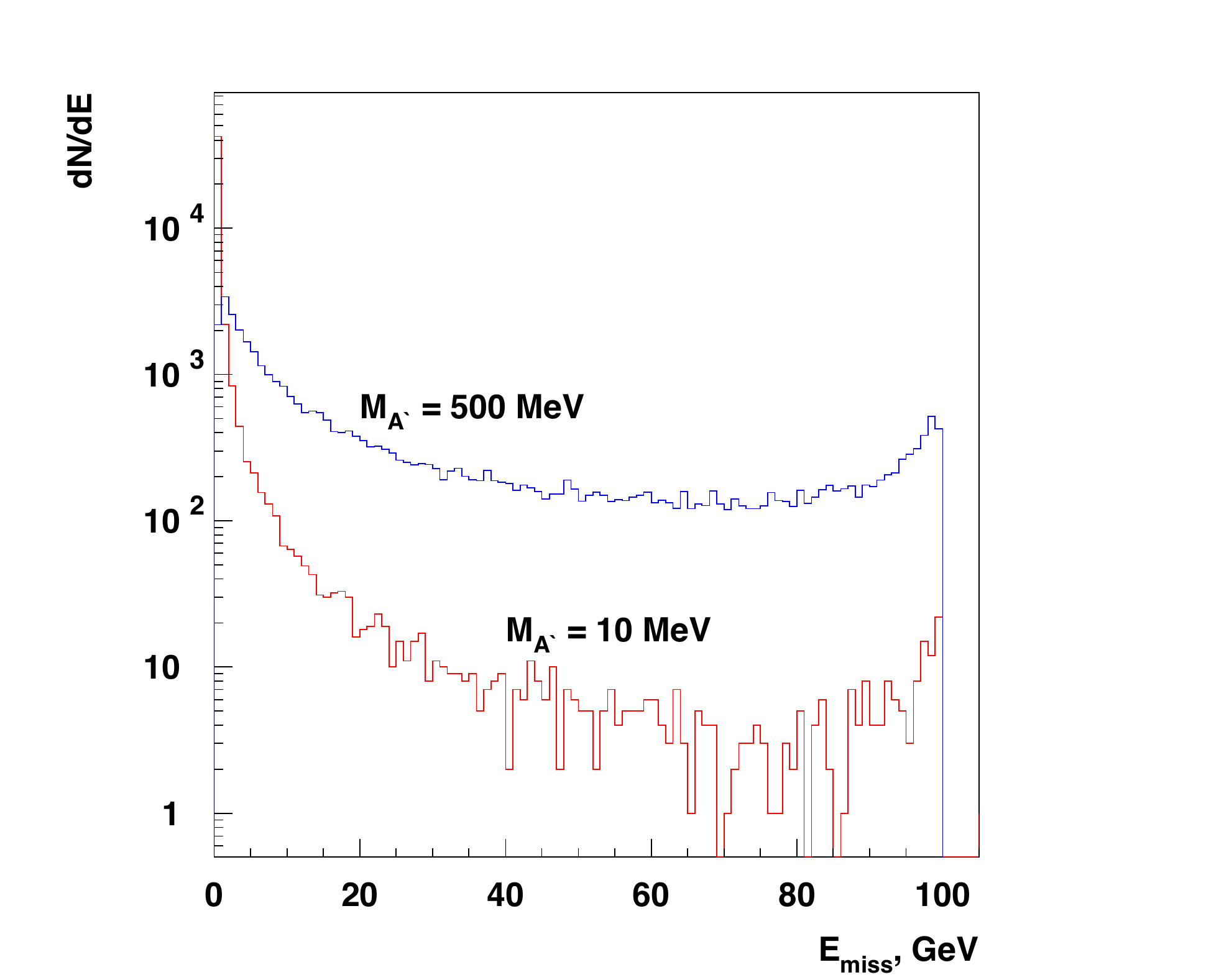}
\caption{ The $A'$ emission spectrum from 100 GeV  electron beam interactions in the Pb target
 calculated for  $m_{A'}=10$~MeV and 
$m_{A'}=500$~MeV. The spectra are normalized to about the same number of events.
\label{AspectraForShower}}
\end{center}
\end{figure}  

\section{The $A'$ production and spectra}

The Lagrangian of the SM is extended by 
the dark sector in the following way:
\begin{eqnarray}
\mathcal{L} = \mathcal{L}_{SM}  -\frac14 F'_{\mu\nu}F'^{\mu\nu}+
\frac{\epsilon}{2}F'_{\mu\nu}F^{\mu\nu}+  \frac{m_{A'}^2}{2}A'_\mu A'^\mu \nonumber \\
+ i \bar{\chi}\gamma^\mu \partial_\mu \chi- 
m_\chi \bar{\chi} \chi -
e_D \bar{\chi}\gamma^\mu A'_\mu \chi, 
\label{DarkSectorLagrangian}
\end{eqnarray}
where $A'_\mu$ is massive vector field of spontaneously broken 
$U'(1)$ gauge group,
$F'_{\mu\nu} = \partial_\mu A'_\mu-\partial_\nu A'_\mu,$
and $\epsilon$ is parameter of photon-paraphoton kinetic mixing. Here, we consider as an
example the Dirac spinor fields $\chi$ which are treated as Dark 
Matter fermions  coupled to 
$A'_{\mu}$ by  dark portal 
  coupling constant  $e_D$.   The mixing term $\frac{\epsilon}{2}F'_{\mu\nu}F^{\mu\nu}$
 results in the interaction:
\begin{equation}
\mathcal{L}_{int}= \epsilon e A'_{\mu} J^\mu_{em}
\end{equation}
of dark photons with the ordinary matter. 
The decay rates of  $A'\rightarrow \bar{\chi} \chi$ and $A'\rightarrow  e^- e^+$ 
 are given by
 \begin{eqnarray}
\Gamma (A'\rightarrow \bar{\chi}\chi) = \frac{\alpha_D}{3} m_{A'}\bigl(1+
\frac{2m_\chi^2}{M_{A'}^2}\bigr)\sqrt{1-\frac{4m_\chi^2}{M_{A'}^2}}, \qquad \nonumber \\  
\Gamma (A'\rightarrow  e^- e^+) = \frac{\alpha_{QED} \epsilon^2}{3} m_{A'}\bigl(1+
\frac{2m_e^2}{M_{A'}^2}\bigr)\sqrt{1-\frac{4m_e^2}{M_{A'}^2}}.
\end{eqnarray}
We suppose that dark matter invisible decay mode is predominant, i.e. 
$\Gamma (A'\rightarrow \bar{\chi}\chi)/\Gamma_{tot} \simeq 1$. 
This means that the $A'$ lepton decay channel is suppresed,
$\Gamma (A'\rightarrow \bar{\chi}\chi) \gg \Gamma (A'\rightarrow  e^- e^+)$.

\begin{figure*}[tbh!]
\includegraphics[width=0.9\textwidth]{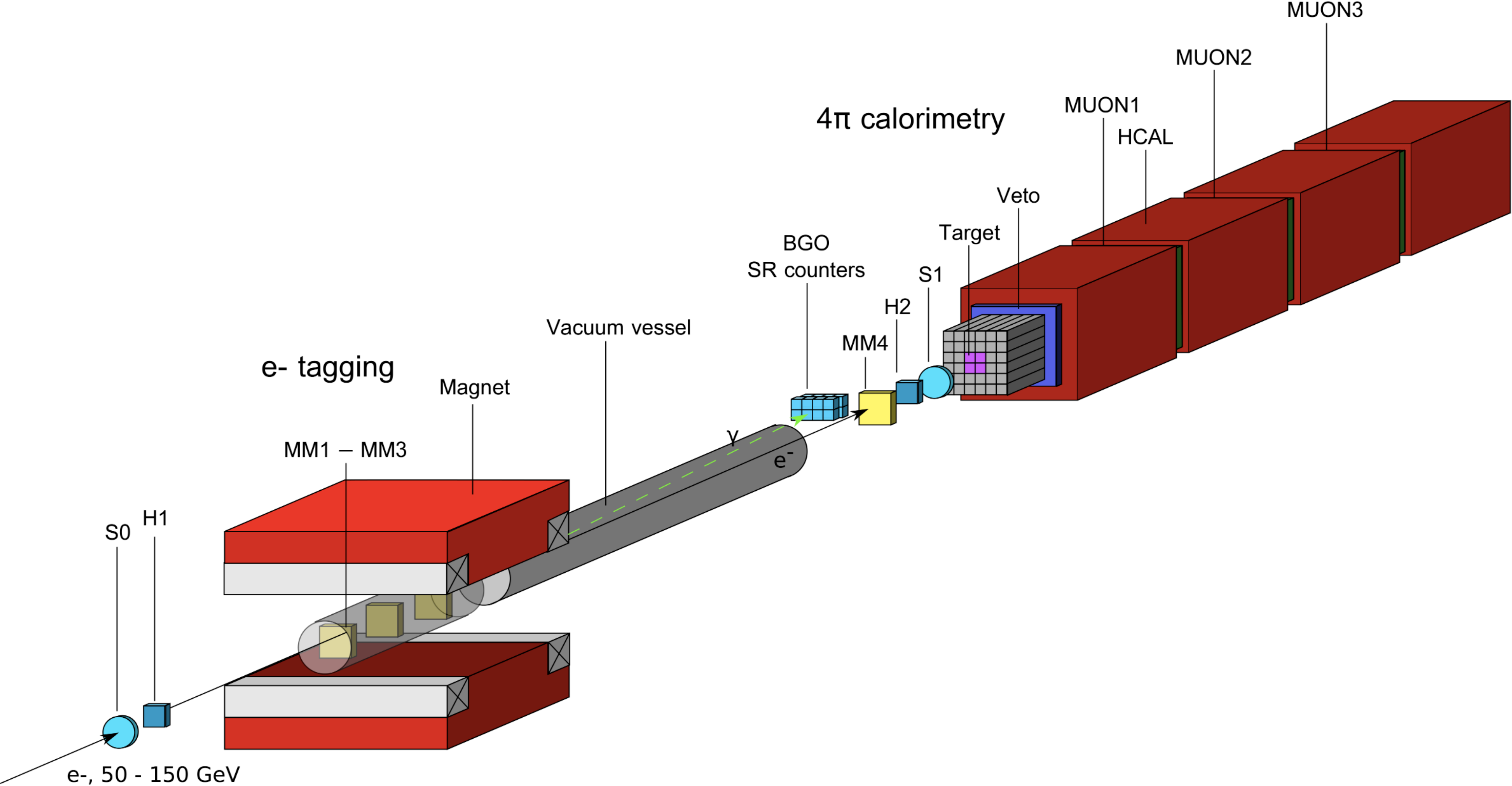}
\caption{Schematic illustration of the setup to search for $\ainv$  decays with 50-150 GeV e$^-$ at H4 beam. The incident electron energy absorption in the ECAL is accompanied by the emission  of  bremsstrahlung $A'$s in the reaction  $eZ\rightarrow eZ A'$ of electrons scattering on nuclei, see Fig.~\ref{diagr}. The part of the primary beam energy is deposited in the ECAL, while  the  remaining fraction of the total energy is  transmitted by the decay dark matter particles   through the rest of the detector resulting in the missing energy   signature in the detector. See text.}
 \label{setup}
\end{figure*} 

We consider the high-energy electron beam absorption in the active target as a source of $A'$s.
In this case dark photons can be produced in the bremsstrahlung off nuclei due to the $\gamma - A'$ mixing (see Fig.~\ref{diagr})
and subsequently decay invisibly ($\ainv$):
\begin{equation}
 e^- Z \rightarrow e^- Z A',~\ainv 
\label{reaction}
\end{equation}
The  $A'$-production cross section in this process 
was calculated \cite{Bjorken:2009mm} in the 
Weizs\"{a}cker-Williams (WW) approximation \cite{Tsai:1986tx}, namely 
\begin{eqnarray}
\frac{d\sigma}{dx \, d \cos \theta_{A'}}= \frac{8
  Z^2\alpha_{QED}^3\epsilon^2 E_0^2 \,
  x}{U^2}\frac{\chi}{Z^2} \nonumber \\
  \left[(1-x+x^2/2)-\frac{x\,(1-x)m_{A'}^2
    E_0^2\,x\,\theta_{A'}^2}{U^2} \right],
    \label{ecrosssection} 
\end{eqnarray}
where $E_0$ is the  energy of incoming electron,  
$E_{A'} $ is the energy of $A'$, $E_{A'} =   x
E_0$,   $\theta_{A'}$ is the angle in the 
lab frame between the emitted
$A'$ and the incoming electron, $Z$ is the atomic 
number of nucleus ($Z = 82$ for lead).
The function 
$U=U(m_{A'},E_0,Z,A)$ 
which determines the virtuality of intermediate 
electron has the following form:
\begin{equation}
U=E^2_0\,x\,\theta_{A'}^2+m_{A'}^2\,
\frac{1-x}{x}+m_e^2\,x. 
\end{equation} 
The effective flux of photons, 
$\zeta = \zeta (m_{A'}, E_0, Z, A)$ is defined as follows:
\begin{equation}
\zeta = \int\limits_{t_{min}}^{t_{max}}\!dt\,\frac{t-t_{min}}{t^2}\,G_2(t),
\label{tinegration}
\end{equation}
where $t = -q^2$, $|\vec{q}|= U/(2E_0 (1-x))$,  $t_{min}\simeq |\vec{q}|^2$,   $t_{max}=m_{A'}^2 $ and
$G_2(t)=G_{2,el}(t)+G_{2,in}(t)$ is the sum of 
elastic and inelastic electric form 
factor (for details see e.g. 
Ref.~\cite{Bjorken:2009mm} 
and references therein). In the numerical integration 
(\ref{tinegration}) we neglect $x$- and
$\theta_{A'}$-dependences of  $t_{min}$.

Several additional remarks should be made.
First, the approximation of collinear $A'$
emission is justified for the benchmark points, $m_{A'}\lsim 1$ 
GeV and  $E_{0}\lsim 100$ GeV, when
$  m_{A'}/E_0 \ll 1$ (see 
Ref.~\cite{Bjorken:2009mm} for details). 
Second, one can perform the cross-section 
(\ref{ecrosssection})
integration over $x$ and $\theta_{A'}$,
\begin{equation}
\sigma_{tot} \simeq \frac{4}{3}
\frac{\alpha^3 \epsilon^2 \, \zeta}{m_{A'}^2} \log (\delta^{-1}),
\label{TotalCrossSectionApr}
\end{equation}
where $\delta=\mbox{max}(m_{A'}^2/E_0^2, m_e^2/m_{A'}^2)$ is 
the infrared (IR) cut-off of the cross-section, which regulates
either soft intermediate electron singularity or validation of WW 
approximation \cite{Bjorken:2009mm}.  

In order to determine the acceptance of the experiment
we perform the signal Monte Carlo simulation.
We simulate the electromagnetic shower
development in the ECAL (See, Section V) with {\tt  GEANT4} using the following steps
\begin{enumerate}[(i)]
\item calculate the total and differential cross-sections
of the $A'$ bremsstrahlung production (\ref{ecrosssection}) as
a function of the electron energy $E_0$,
\item at each step of an electron propagation in the lead converters of the ECAL,
 the emission of the $A'$ is randomly generated,
\item if the emission is accepted, then we generate values
of $x$, $\cos \theta$, and the azimuthal angle $\phi_{A'}$,
\item finally, the 4-momentum of the recoil electron is calculated.
\end{enumerate}

In Fig.~\ref{AspectraForShower} an example of the $A'$ energy distributions calculated for  
masses $m_{A'}=10$ MeV and $m_{A'}=500$ MeV are shown. Note that these distributions represent
also the missing energy spectra in the detector.

\section{The detector}

The  $A'$ production is a  rare  event. For the interesting  parameter range it is  expected to occur with a  rate $\lesssim  10^{-9}$ with respect to the  ordinary photon production rate. Hence, its observation represents a challenge for the detector
design and performance. 

The  experimental setup  specifically designed to search for the $A'$ production in the reaction 
\eqref{reaction} of high-energy electron scattering off  nuclei in a high density target $T$ is schematically shown in Fig.~\ref{setup}.  The experiment  employs the upgraded H4 electron
beam line  at the CERN SPS described in details in 
 Ref.\cite{beam}. The  beam is designed to transport the electrons with the maximal 
 intensity $\simeq (3-4)\cdot 10^6$ per SPS spill in the  momentum range between 50 and 150 GeV/c that could be produced by the primary proton beam of 450 GeV/c with the intensity up to a few 10$^{12}$ protons on target.
The electrons are produced by protons impinging  on a primary beryllium target and transported to the detector inside the evacuated beam-line tuned to an adjustable  beam momentum. 
The hadron contamination in the electron beam is  $\pi/e^- \lesssim 10^{-2}$ and the size of the beam at the detector position  is of the order of a few cm$^2$.

The detector shown in Fig.~\ref{setup} utilizes  upstream magnetic spectrometers (MS)
consisting of  dipole magnets and a low-material-budget tracker, which is a set of Micromegas  chambers , MM1-MM4,  allowing the  reconstruction and  precise measurements of momenta for incident electrons \cite{Banerjee:2015eno}. It also uses the scintillating counters S0, S1 and hodoscopes H1 and H2 to define the primary beam, and  the active target $T$,   which is the central part of the high-efficiency hodoscopic   electromagnetic calorimeter (ECAL) used for the accurate measurement of the the recoil electron energy from the reaction \eqref{reaction}. Downstream the target the detector
 is equipped with  high-efficiency forward veto counter V, and a massive, completely hermetic hadronic calorimeter (HCAL).   Three straw-tubes chambers, MUON1-MUON3,  located between the HCAL 
 modules are  used for the final-state muon(s) identification. The modules serve  as a dump to completely absorb  and detect the energy  of hadronic secondaries produced in the electron  interactions $e^- A \to anything$ in the target. 
    In order to suppress  backgrounds caused by the detection inefficiency  the HCAL must be longitudinally completely hermetic \cite{Gninenko:2013rka, Andreas:2013lya}. To enhance its hermeticity, the HCAL thickness is chosen to be $\simeq 30 ~\lambda_{int}$ (nuclear interaction lengths).  
The 15 m long vacuum vessel between the magnet and the ECAL is installed to avoid absorption of the synchrotron radiation photons detected at the downstream end of the vessel by the array of BGO crystals for the effective tagging of the incoming beam electrons \cite{Gninenko:2013rka}. 
\begin{figure*}[tbh!!]
\includegraphics[width=0.55\textwidth]{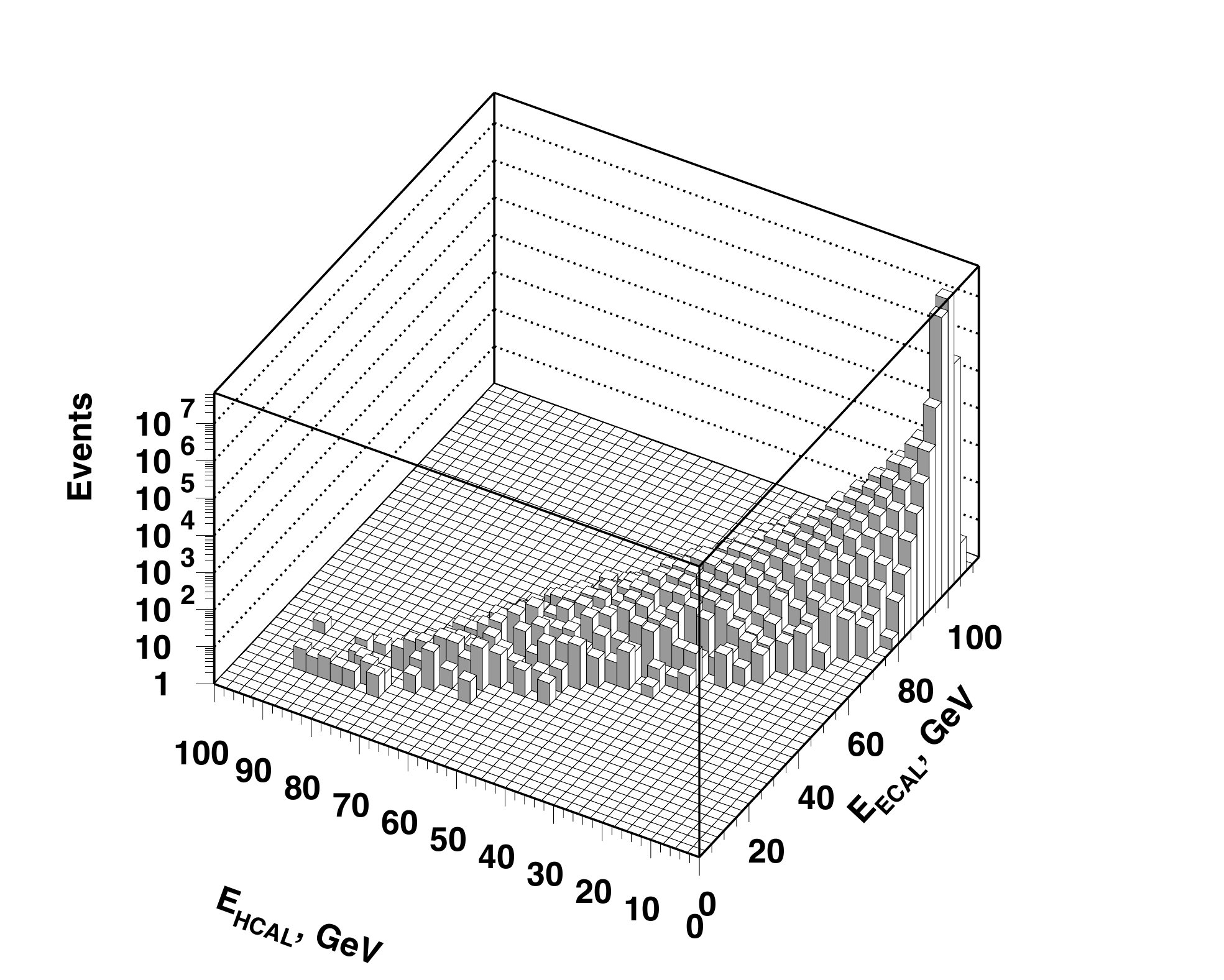}
\hspace{-2.cm}{\includegraphics[width=0.55\textwidth]{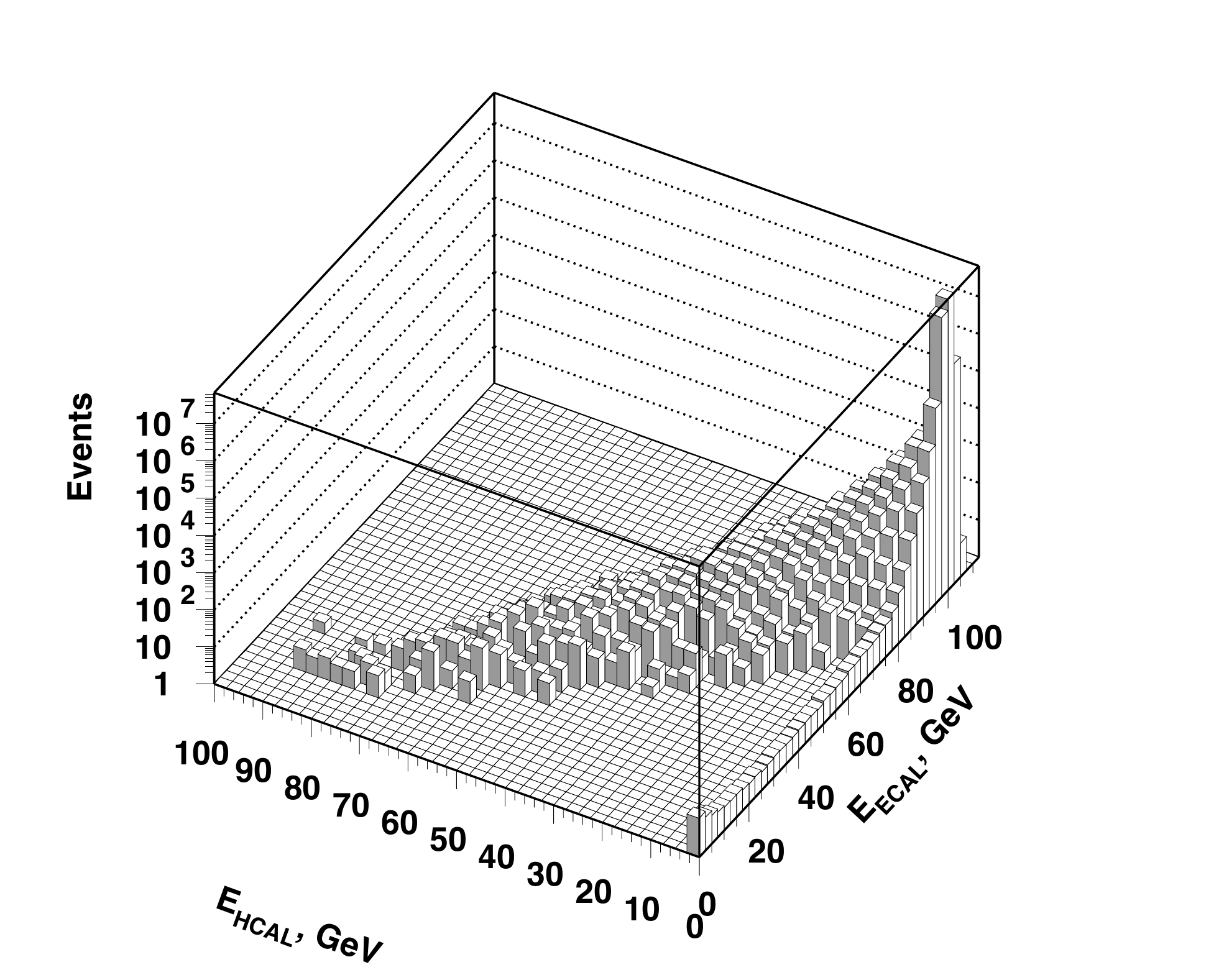}}
\caption{  Expected distributions of events in the ($E_{ECAL}$;$E_{HCAL}$) plane from the  SM interactions induced by the 100 GeV $e^-$'s in the ECAL target (left plot) and from the same reactions plus  
the $A'$ emission in the process \eqref{reaction}(right  plot). Every event in the left plot satisfies within the
uncertainties the constraint $E_{ECAL}+E_{HCAL} = E_0$. In the right plot the events from the region 
  $0 \lesssim E_{ECAL} \lesssim 80$ GeV, $E_{HCAL}\lesssim 1$ GeV  have $E_{ECAL}+E_{HCAL} < E_0$ due to the loss
 of a significant fraction of energy which is carried away by $A'$s. The $A'$ energy  spectrum is calculated for the mixing value $\epsilon \simeq 10^{-2}$ and  mass $M_{A'} = 50$  MeV.}
\label{sig-bckg}
\end{figure*}   
\section{missing energy signature of signal events}

 The method of the search is the following \cite{Gninenko:2013rka}.  The reaction~\eqref{reaction} typically occurs in the first few radiation length ($X_0$) of the ECAL. The part of the primary beam energy is deposited in the ECAL, while  the  remaining fraction is  transmitted by  the decay particles $\chi$  through the rest of the detector. As the $\chi$s are very weekly interacting particles, they  penetrate the ECAL,  veto V and the HCAL  without interactions  resulting in the missing energy   signature in the detector, see Fig.~\ref{setup}.
The occurrence of $\ainv$ decays would appear as an excess of events with single e-m showers in the ECAL,  and zero energy deposition in the rest of the detector, above those expected from the background sources. The signal candidate events have the signature:  
\begin{equation}
S_{A'} = {\rm H1 \times H2 \times ECAL(E_{ECAL}< E_0) \times \overline{ V \times HCAL}},
\label{signinv}
\end{equation}
and should satisfy the following selection criteria:  
\begin{enumerate}[(i)]
\item The momentum of the incoming particle track  should correspond to the beam momentum.
\item The starting point of the (e-m) shower in the ECAL should be localized  within a few first $X_0$s.  
\item The lateral and longitudinal shape of the shower in the ECAL is consistent with the  one expected for the signal shower. The fraction of the total  energy deposition in the ECAL is $f\lesssim 0.5-0.7$.  where $E_0$  is the benchmark electron beam energy, 
 $E_0=100$ GeV.  This implies  the selection condition  for the recoil electron  $E_{e}'<50$ GeV. Therefore,  the missing energy $E_{mis}=E_{A'} =E_0 - E_{ECAL}$ should be  $E^e_{mis}=E_{A'}> E_{0}/2$.
\item No energy deposition in  the V and HCAL.
\end{enumerate}
\begin{figure*}[tbh!!]
\includegraphics[width=0.55\textwidth]{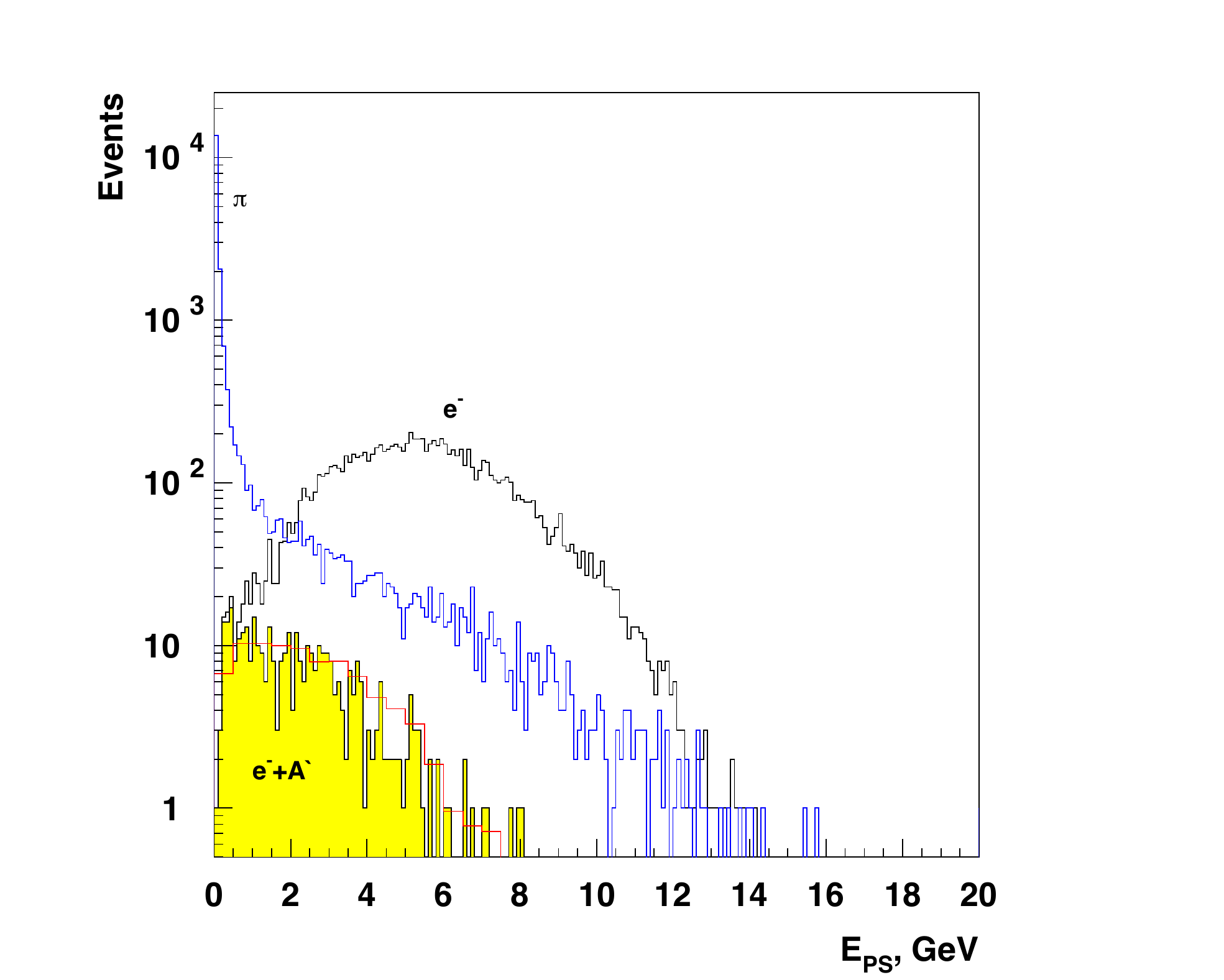}
\hspace{-2.cm}{\includegraphics[width=0.55\textwidth]{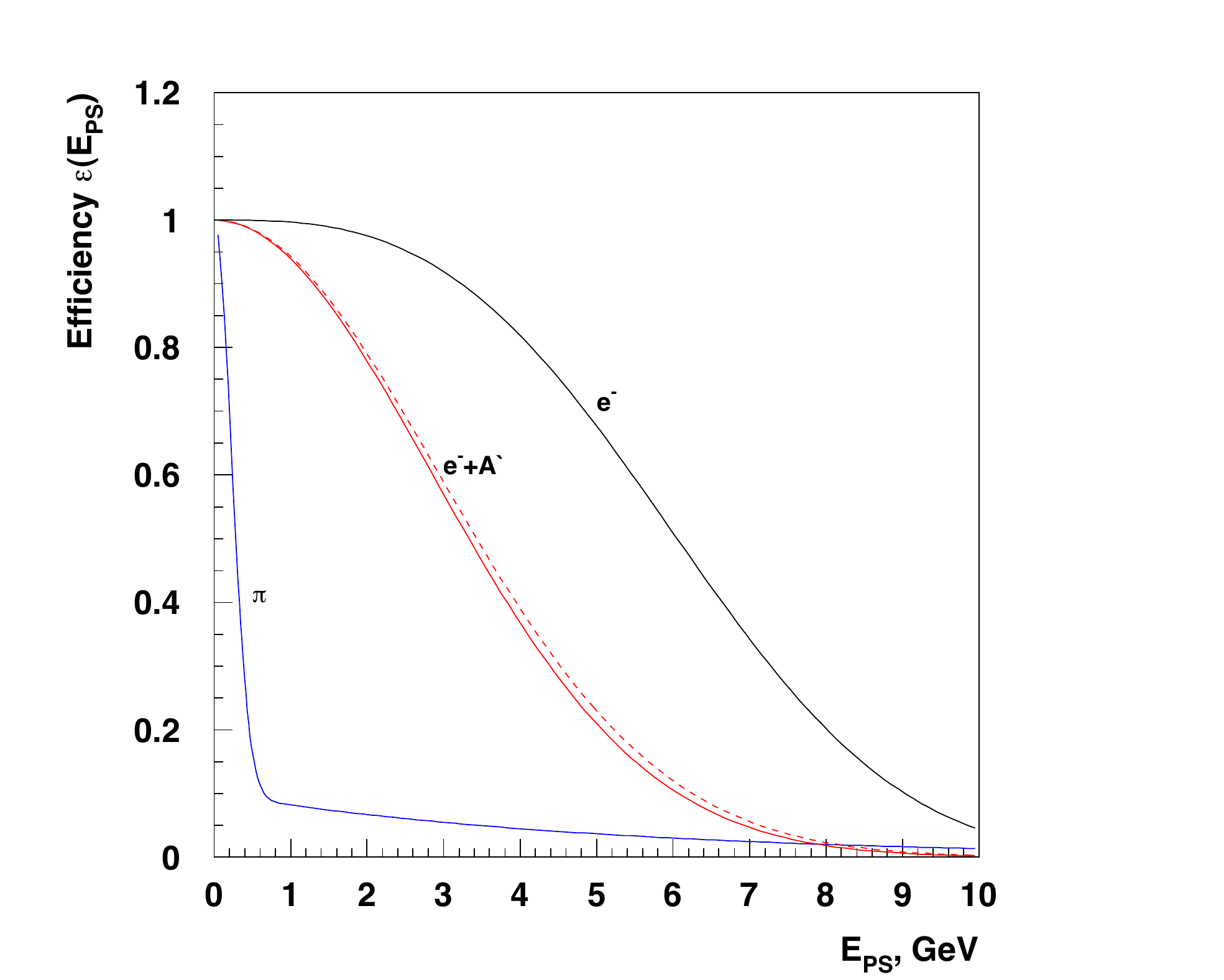}}
\caption{  Expected distributions of the energy deposited in the ECAL preshower from 100 GeV pions (blue), electrons (black) and signal  events for $M_{A'} = 50$ (shaded) and $= 200$  MeV (red)  (lhs plot). The  energy  spectrum of $A'$s emitted  in the reaction \eqref{reaction} is calculated for the  mixing strength 
$\epsilon \lesssim  1 $ and $E_{miss}\gtrsim 0.5 E_0$. The rhs plot shows the pion, electron and signal efficiency, $M_{A'} = 50$ (red)  and 200 (red dashed)   MeV, as a function of threshold on the $E_{PS}$ value. }
\label{prs}
\end{figure*}   

In Fig. \ref{sig-bckg} the expected distributions of events in the ($E_{ECAL}$;$E_{HCAL}$) plane 
from the  SM interactions induced by the 100 GeV $e^-$'s in the ECAL target (left plot) and from the same reactions plus  the $A'$ emission in the process \eqref{reaction}(right  plot). The only event selection criterium used is the requirement of no simultaneous signals in the muon counters MUON2 and MUON3 and energy deposition in the last two downstream HCAL modules. Such signature  give evidence for the presence of minimum ionizing particles (MIP), presumably muons, in the final state which typically originate from the 
$\pi, K \to \mu \nu$ decays in-flight resulting in the missing energy in the event  due to emission  of neutrinos.  
One can see that the experimental signature of the 
$A'$ production in the reaction \eqref{reaction} is an event with the missing energy 
$E_{miss}\gtrsim E_0 - E^{th}_{ECAL}$ from the region
$0\lesssim E_{ECAL} \lesssim E^{th}_{ECAL}$ and  $0\lesssim E_{HCAL} \lesssim E^{th}_{HCAL}$.
The typical values for the ECAL and HCAL threshold energies are expected to be 
 $E^{th}_{ECAL} \simeq 50$ GeV, i.e. $E_{miss} > 50$ GeV,  and $E^{th}_{HCAL} \simeq 0.3$ GeV, respectively. 
The events in this region   are supposed to be from the reaction \eqref{reaction}
as a large fraction of the primary beam energy is  carried away by the $A'$,  those  spectrum shown in Fig.~\ref{sig-bckg} for  $M_{A'} = 50$  MeV, and mixing strength $\epsilon \simeq 10^{-2}$.
For the ECAL, the value of $E^{th}_{ECAL}$ is defined by the shape of the low energy tail 
of  the ECAL response function to the monochromatic electron  beam. This tail is mostly due to i) the longitudinal fluctuations of the e-m shower development and corresponding leak 
energy, and ii) electroproduction of hadrons by primary  electrons in the target. 
The $E^{th}_{HCAL}$ value is defined mostly by the noise level of the HCAL electronics, energy leak from the ECAL, and pileup events, see Section V.    
 
     The distributions shown in Fig. \ref{sig-bckg}  are obtained with $\sim 10^8$  simulated with GEANT4 events. Due to the small coupling strength of the $A'$ the reaction \eqref{reaction} occurs typically 
with the rate $\lesssim 10^{-9}$ per the incoming electron interaction. To  study the SM distribution and  background (see Section V)  at this level would require the simulation of a very large number of events resulting in a prohibitively large amount of computation time. Consequently, only $\lesssim 10\%$ of the required statistics for the SM  reactions were simulated.

\begin{figure*}[tbh!!]
\includegraphics[width=0.55\textwidth]{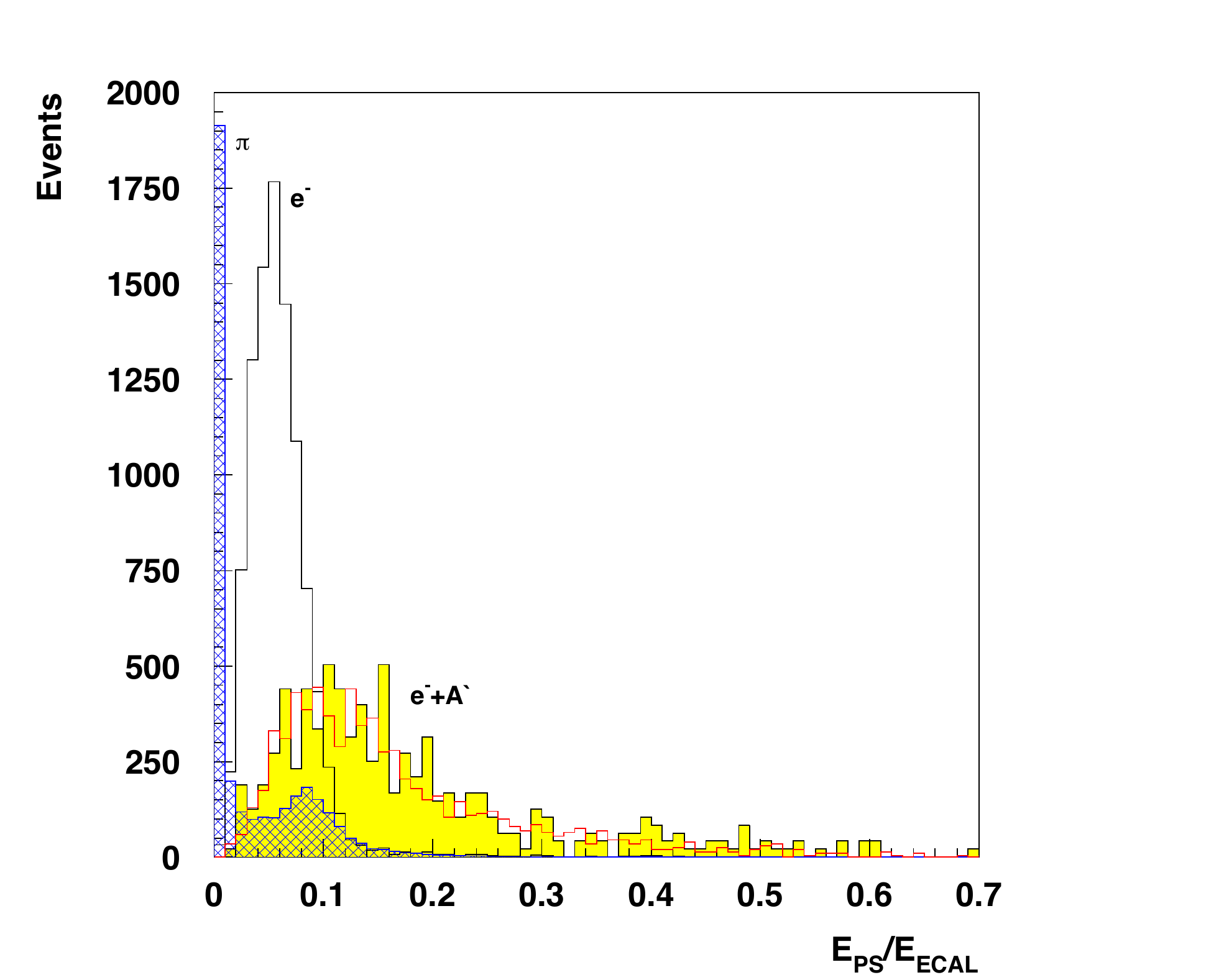}
\hspace{-2.cm}{\includegraphics[width=0.55\textwidth]{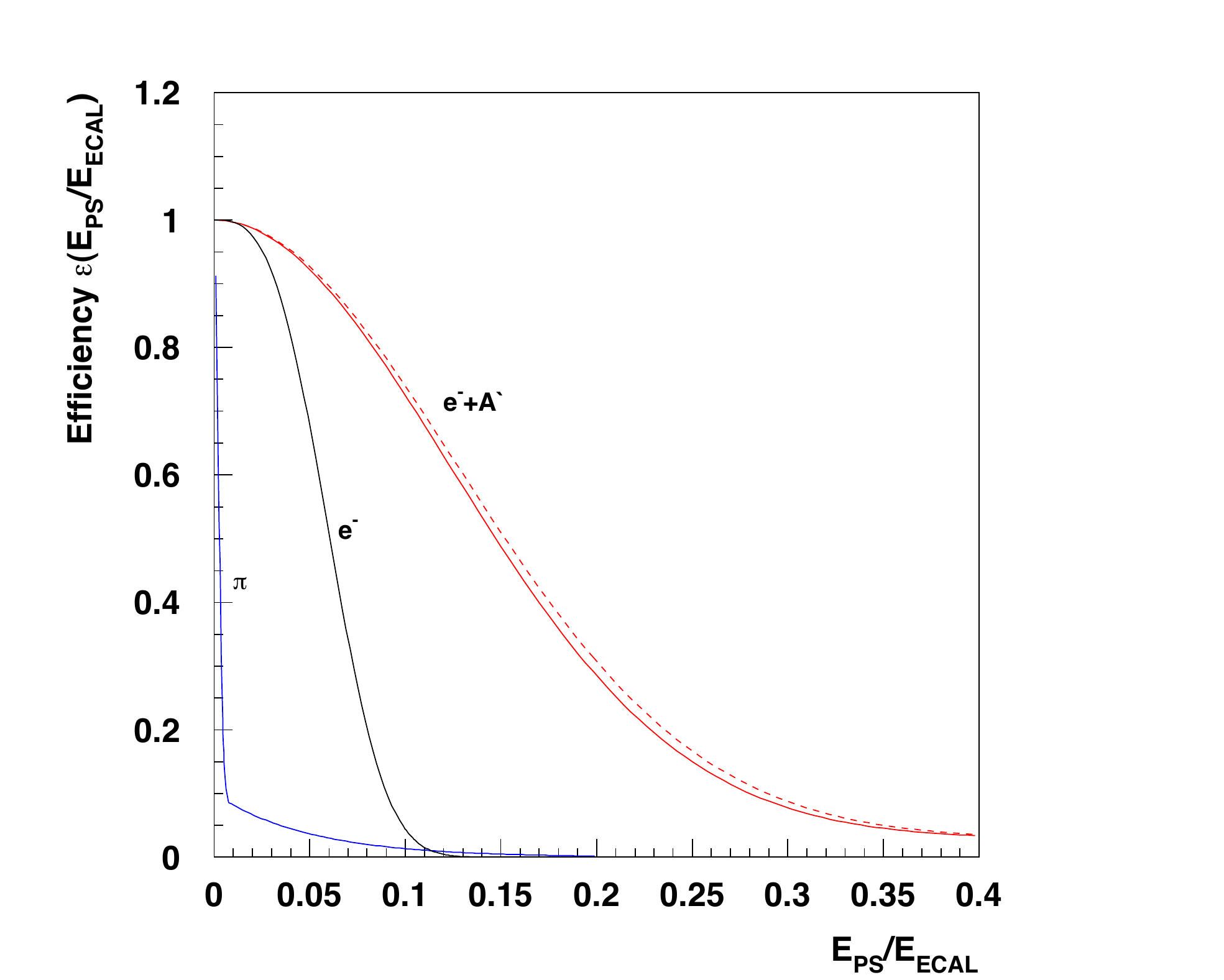}}
\caption{  Expected distributions of the $E_{PS}/E_{ECAL}$ ratio of energy deposited in the ECAL preshower to the total energy deposited in the ECAL (lhs plot)  from 100 GeV pions (blue), electrons (black) and signal  events for $M_{A'} = 50$ (shaded) and $= 200$  MeV (red). The  energy  spectrum of $A'$s emitted  in the reaction \eqref{reaction} is calculated for the  mixing strength 
$\epsilon \lesssim  1 $ and $E_{miss}\gtrsim 0.5 E_0$. The rhs plot shows the pion, electron and signal efficiency, $M_{A'} = 50$ (red)  and 200 (red dashed)   MeV, as a function of threshold on the $E_{PS}/E_{ECAL}$ value.}
\label{ratio}
\end{figure*}   

\section{The ECAL response to the signal events}

The use of the ECAL is twofold. On the one hand, it serves as an active target to measure 
the total energy deposition in the beam dump. On the other hand,
 it has longitudinal and lateral granularity allowing additional  
suppression of the hadronic background by studying the shower shape.
Simulations performed with GEANT4  show that by using the electromagnetic and hadronic shower profiles in the calorimeters, both lateral and longitudinal,
it is possible to further improve the $e/\pi$ rejection by a factor of 5-10.
Longitudinally the ECAL is subdivided in two parts:  preshower and absorption part. The preshower has 4 radiation lengths of lead
and plays an important role in the hadron background rejection obtainable with the ECAL. Hadron rejection is ultimately limited by such processes as charge exchange ($\pi^\pm + N \to n \pi^0 + N' $) where most of the energy of the charged pion goes to one or more neutral pions. The $\pi^0$s immediately decays into photons starting a cascade shower which is indistinguishable from the electron-initiated shower. Thus,  charge exchange interactions
of the beam pions  occurring near the front of the ECAL array and accompanied by a poor detection of the rest of the final state cannot be separated from the reaction \eqref{reaction} \cite{Gninenko:2013rka}.
The additional suppression of such processes can be provided by using the lead as the calorimeter passive material
(it has a smaller number of interaction lengths per radiation length) by the requirement of the early development of the shower.
\begin{figure*}[tbh!!]
\includegraphics[width=0.55\textwidth]{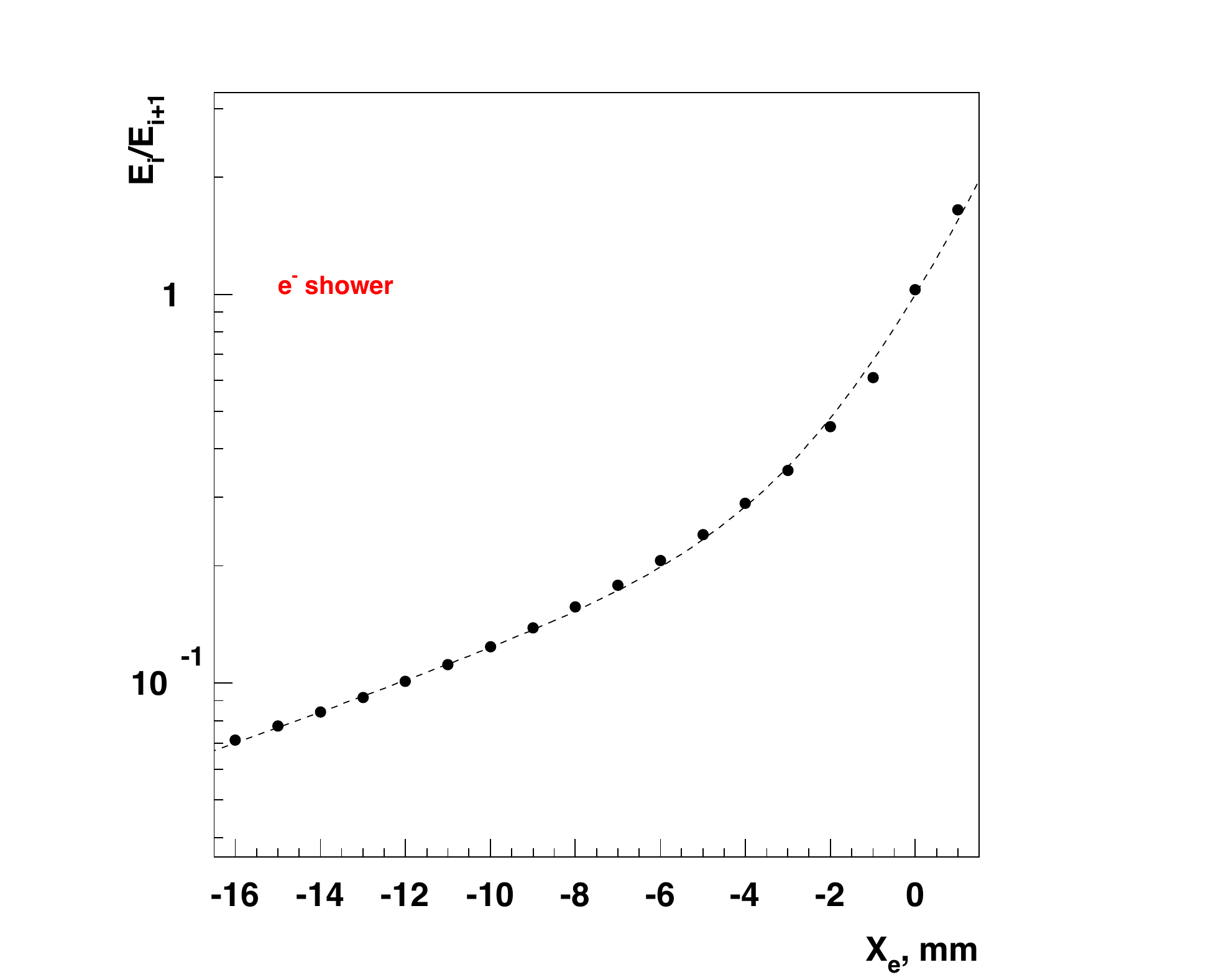}
\hspace{-2.cm}{\includegraphics[width=0.55\textwidth]{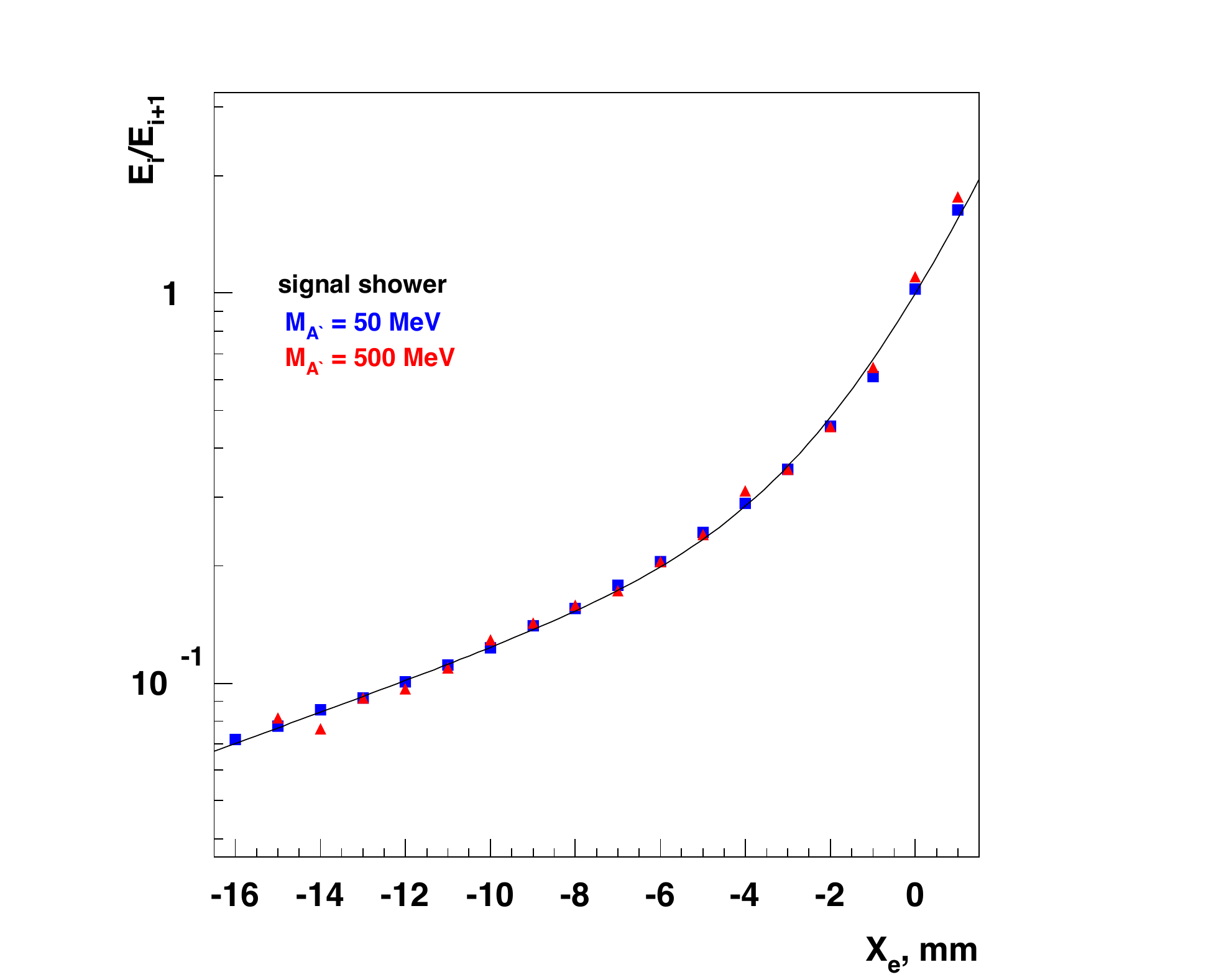}}
\caption{Expected dependence of the ratio $E_i/E_{i+1}$ of the average energies deposited in two  adjacent cells on electron coordinate $X_e$ for the electromagnetic showers induced by the beam electrons without (left plot) and with (right  plot)  the $A'$ emission from the reaction \eqref{reaction}. Shown are the simulated  values, the curves have been calculated for the lateral shower profile of Eq.(\ref{profile}), see text.} 
   \label{profile}
\end{figure*}
The $A'$ events are supposed to be reconstructed in the ECAL as electromagnetic showers.
Therefore, the question arises to what  extend the properties of the electromagnetic shower
in the ECAL from the reaction  \eqref{reaction} are identical to the properties of the ordinary shower induced by an electron with the same energy deposition in the ECAL. For example, one could suggest that the emission  of a high-energy $A'$ could make the residual electromagnetic shower development slightly asymmetric resulting in modification of the lateral shower profile exceeding the ordinary shower fluctuations.
This could  change the selection efficiency of the cuts iii) of Section IV. 
 To answer this question we have compared the lateral and longitudinal electromagnetic shower profiles in the ECAL for ordinary and signal electromagnetic showers  induced by the reaction \eqref{reaction}. In this study  the shashlik ECAL used in simulations has  the following characteristics:   
\begin{enumerate}[(i)] 
\item  It is a matrix of 6x6 cells, each with dimensions $38.2\times 38.2 \times 490 $ mm$^3$.
\item Each cell is (1.50 mm Pb + 1.50 mm Sc) x 150 layers or 40 radiation length ($X_0$).
\item Each cell is longitudinally subdivided into two parts:  preshower section (PS)  of 4 $X_0$ and the main ECAL of 36 $X_0$. 
\item The simulated energy resolution  is  $\sigma E/E \simeq 9\%/\sqrt{E(GeV)}+0.7$
\end{enumerate}

\subsection{Longitudinal shower development} 

One of the sources of background is expected from  hadron interactions in the  ECAL that could 
mimic the signal \cite{Gninenko:2013rka}. 
  The electron-hadron separation in this case can be improved if we measure the electromagnetic shower development at an early stage by using the ECAL preshower section. 
Then the question arises how identical are  the longitudinal development of showers induced by  the signal reaction \eqref{reaction} and by an ordinary electron and how the applied hadron rejection cuts affect the signal efficiency.      
In this section, we take a  step toward answering this question. We examine  the qualitative features associated with the longitudinal distributions of deposited energy by showers induced by pions, electrons and signal events,  and assess to what extent these features are affected by dark-photon emission for the signal events.  
\begin{figure}[tbh!]
\begin{center}
\includegraphics[width=0.5\textwidth]{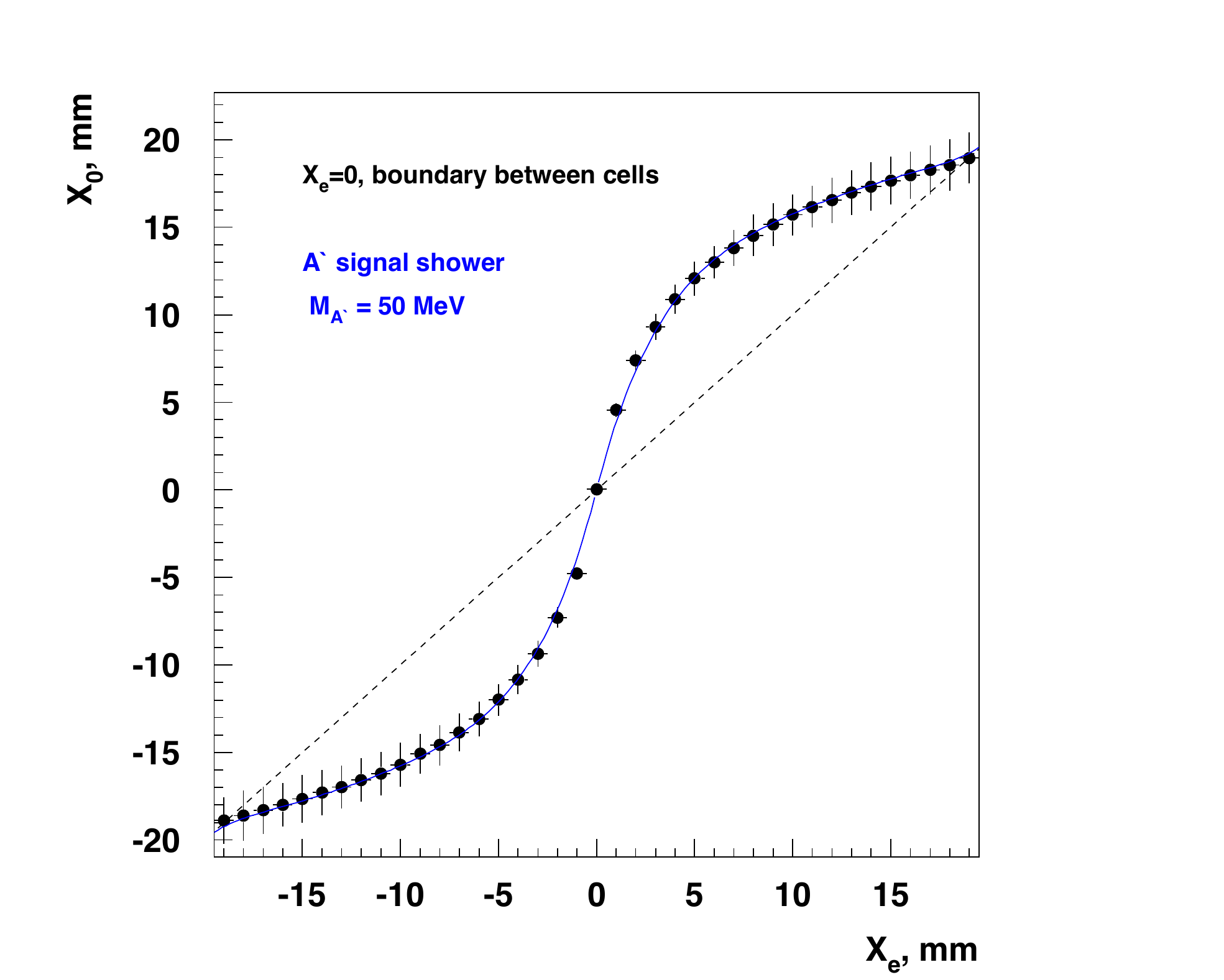}
\caption{ The calculated with Eq.(\ref{cog}) coordinates of the shower centre-of-gravity $X_0$ at different  positions of the true coordinate $X_e$ of incoming electrons. The position $X_e = 0$ corresponds to the cell centre. Dots show the reconstructed  values for $X_0$, the error bars represents  uncertainties ($\sigma_X$) in the coordinate $X_c$ reconstruction. The curve has been calculated with Eq.(\ref{shpr}) for $M_{A'} = 50$ MeV .}
\label{xcoord}
\end{center}
\end{figure}  
\begin{figure*}[tbh!]
\begin{center}
\includegraphics[width=1.\textwidth]{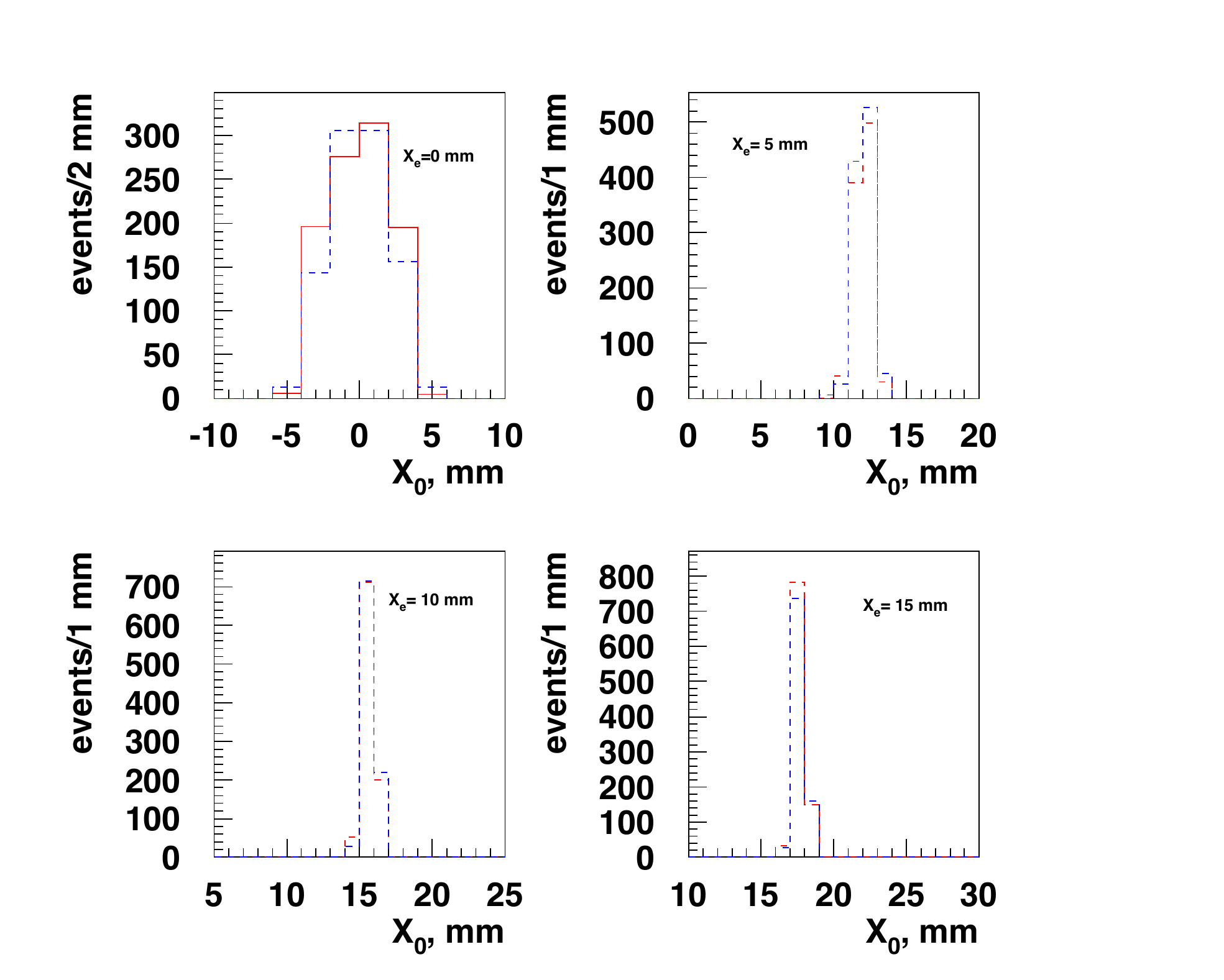}
\caption{ The calculated with Eqs.(\ref{shpr}, \ref{cog}) coordinates of the shower centre-of-gravity $X_0$ at different  positions of the true coordinate $X_e$ of incoming electrons. The position $X_e = 0$ corresponds to the cell centre. Dots show the reconstructed  values for $X_0$, the error bars represents  uncertainties ($\sigma_X$) in the coordinate $X_e$ reconstruction. The curve has been calculated with Eq.(\ref{shpr}).}
\label{xcoord1}
\end{center}
\end{figure*}   

  We  use variable $r=E_{PS}/E_{tot}$ - the ratio of the energy deposit in the PS to the total energy deposit $E_{tot}$ - to  evaluate the pion rejection factors at given electron and signal efficiencies. 
   The distribution of energy $E_{PS}$ deposited in PS and the $r$ ratios  for 100 GeV showers induced by the pions, electrons, and signal events is shown in the lhs of
  Fig.~\ref{prs} and  Fig.~\ref{ratio}, respectively.  The electron, pion and signal efficiencies as functions of the threshold on the $E_{PS}$ and $r$ values  are shown in the rhs of Fig.~\ref{prs} and Fig.~\ref{ratio}, respectively. For the signal events the 
  calculations are performed for $E_{miss}> 0.5 E_0$. One can see that for the 
$A'$ case the fluctuations of the  $E_{PS}/E_{tot}$ ratio are significantly large 
then for the electron case: the $r$ value ranges from 0 to $0.6$, while for the 
electron induced events it is in the region $0<r<0.1$.  By comparing spectra, one can also see that 
distributions for signal events are  weakly dependent on the $A'$ mass.
Interestingly, for the same threshold $E_{PS}^{th}$ on the $E_{PS}$ value, the electron efficiency 
$\epsilon_{e}(E_{PS}^{th})$ is higher than the signal one, $\epsilon_{A'}(E_{PS}^{th})$ as shown in Fig.~\ref{prs}. In order to keep  $\epsilon_{A'}(E_{PS}^{th})\gtrsim 0.9$ the threshold should be  $E_{PS}^{th} \lesssim 1 $ GeV. However, for the same threshold $r^{th}$, the situation is opposite, and the signal efficiency is higher compared to the electron one,  $\epsilon_{A'}(r^{th}) > \epsilon_{e}(r^{th})$, as shown in Fig.~\ref{ratio}. This is because 
the emission of the $A'$ with the energy $E_{A'} > 0.5 E_0$ is typically occurs  in the early stages of the electromagnetic shower development. After the $A'$ emission, the residual shower has much lower energy than the primary electron energy,  and thus is also shorter in length. Therefore, larger fraction of its energy is  deposited in the first PS part of the ECAL.

 \subsection{Lateral shower development} 

 Fig.~\ref{profile} shows the simulated  dependence of the average  ratio $E_i/E_{i+1}$ of energies deposited in  two adjacent counters on the electron coordinate $X_e$ for both electron and signal showers for 
 masses $M_{A'} = 50$   and 500   MeV. The coordinate $X_e = 19.1$ mm corresponds to the centre of the (i+ 1)th cell of the ECAL, while $X_e=0$  is the boundary between the ith and (i + 1)th cells. With the obtained dependence of the ratio 
$E_i/E_{i+1}$ on $X_e$ one can define the shower profile $E(X_e)$, which is the energy release as a function of the distance from the shower  axis well described by two  exponential functions:
\begin{eqnarray}
E(x_e) = a_1 exp(-|x_e|/b_1)  \nonumber\\
+ a_2 exp(-|x_e|/b_2)
\label{shpr}
\end{eqnarray}
The fit shown in Fig.~\ref{profile} results in $b_1=2.1\pm 0.3$ mm, $b_1=12.3\pm 1.3$ mm
and $a_1/a_2=0.14 \pm 0.03$ for electron and $b_1=2.1\pm 0.3$ mm, $b_1=12.3\pm 1.3$ mm ($b_1=2.15\pm 0.3$ mm, $b_1=11.9\pm 1.4$ mm) 
and $a_1/a_2=0.14 \pm 0.03$ ($a_1/a_2=0.13 \pm 0.04$) for signal evens with $M_{A'} = 50$ (500) MeV, which are in good agreement with each other for both mass values.  

The simplest method to determine the coordinates of high energy photons and electrons in a granular calorimeter is to measure
the "center of gravity" $X_0$ of the electromagnetic shower induced by them \cite{Akopdzhanov:1976pr}:
\begin{equation}
X_0=2 \Delta \sum_i i E_i /\sum_i E_i,
\label{cog}	
\end{equation}
where $\Delta$ is the half-width of the ECAL cell. 
In Fig.~\ref{xcoord} the calculated with Eqs.(\ref{shpr},\ref{cog}) coordinates of the shower centre-of-gravity $X_0$ at different
positions of the true coordinate $X_e$ of incoming electrons are shown. The position $X_e = 0$ for this case corresponds to the cell centre.
Dots show the reconstructed  values for $X_0$, the error bars represent uncertainties ($\sigma_X$) in the
coordinate $X_c$ reconstruction. The reconstructed $X$-coordinate of the signal e-m showers
for both cases shown in Fig.~\ref{xcoord1} are shifted 
with respect to the true coordinate  of the primary electron $X_e$. The distributions
are found to be very similar to each other. For example, they are practically identical 
  for the beam positioned at the boundary 
 between the cells, where the difference due to transverse shower fluctuations is expected to be most significant. 
The deviation from linearity is due to the two-exponential shape of the 
e-m shower profile in the ECAL calculated with Eq.(\ref{shpr}) for pure electron and signal 
events. One can see that both dependences are very similar. This nonlinearity can be corrected with technique described e.g. in Ref.\cite{Akopdzhanov:1976pr,Davydov:1976vm}.
From Fig.~\ref{xcoord1} we conclude that the shape selection efficiency
for signal events with given X,Y cuts will not differ from the efficiency for pure electrons with the same energy deposition in ECAL.
\begin{figure}[tbh!]
\begin{center}
\includegraphics[width=.5\textwidth]{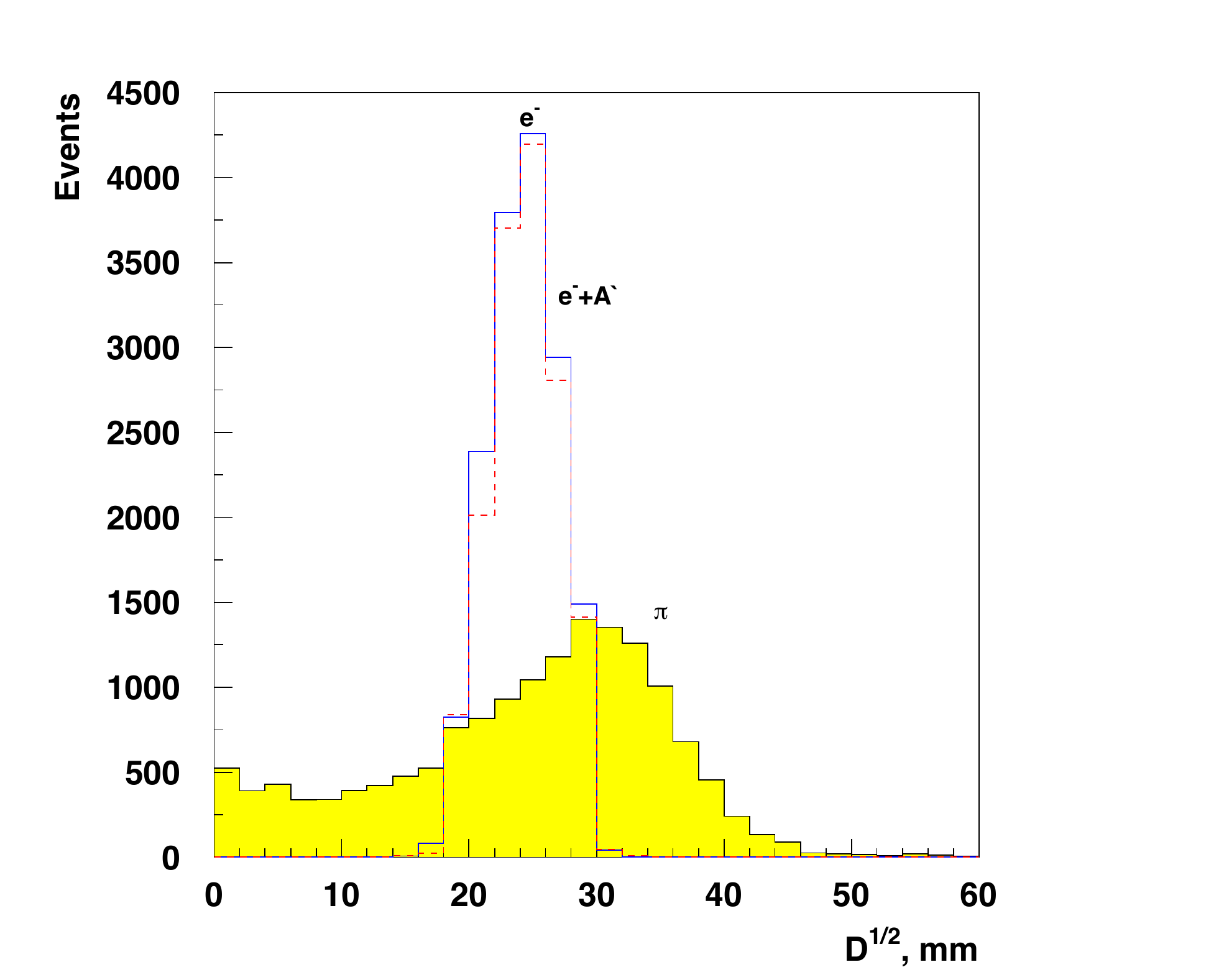}
\caption{ The calculated distribution of the $D^{1/2}$-variable  of Eq.(\ref{disper}) for  showers 
produced in the ECAL by the 100 GeV electron (blue), pions (yellow shaded) and signal events 
with  the $A'$ mass 200 MeV (red dashed). }
\label{hwidth}
\end{center}
\end{figure}  
As discussed previously,   simulations  of the energy response to hadrons show that there is a non-zero probability that the observed energy
deposition, e.g. of a pion is consistent with that of an electron. The lateral shower shape information can also be used to reduce
the probability of primary electron misidentification. As a characteristic for the shower width in the ECAL we have used  its dispersion $D$,  which  can be defined as \cite{Davydov:1976vm}:
\begin{equation}
D = \sum_i  E_i \bigl[(X_i - X_e)^2 +(Y_i-Y_e)^2\bigr]^{1/2}/ \sum_i E_i,
\label{disper}
\end{equation}
where $X_i,~ Y_i$ are the $X,Y$ coordinates of the center of the ith cell.  
The simulated distribution of the $D^{1/2}$-value, representing the "effective radius" of showers induced in the ECAL by the 100 GeV electron, pions and signal events for the $A'$ with mass 200 MeV is shown in 
Fig. \ref{hwidth}.  As one can see  from the figure, electron and signal showers in the ECAL are practically identical, but differ essentially in their widths from  hadronic showers.   By introducing  criterium to select the showers by their dispersion allows one to suppress hadron detection by an additional factor $\simeq 3$, which 
is weakly depends on the $A'$ mass. 
\begin{figure}[tbh!]
\begin{center}
\includegraphics[width=.5\textwidth]{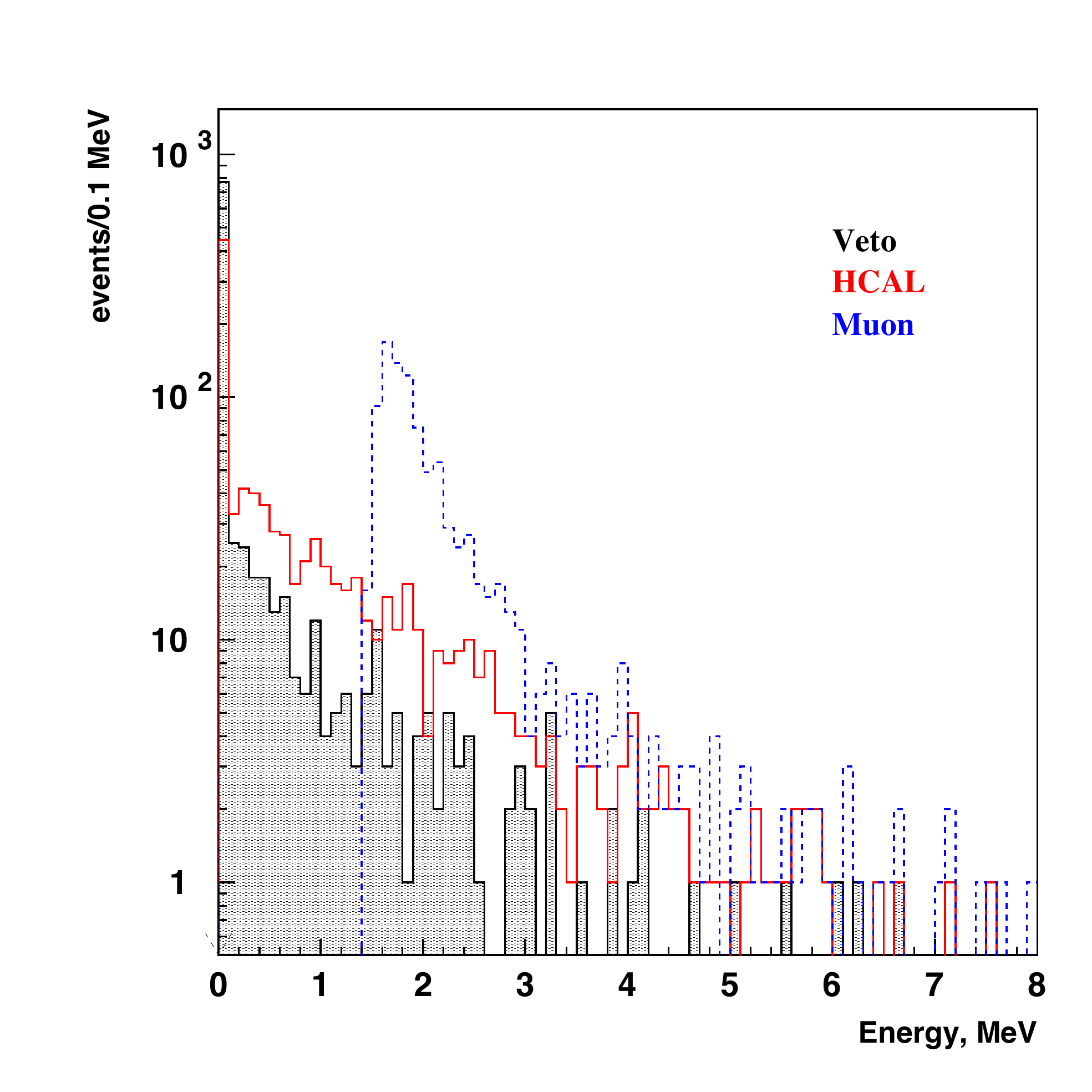}
\caption{ The calculated distribution of energy deposited in the Veto and HCAL
from the reaction \eqref{reaction}. 
The distribution of the energy released in the Veto by a muon is also shown for comparison.}
\label{ecalleak}
\end{center}
\end{figure}

 Finally, the main conclusion of this study is that while  the  properties of electromagnetic showers induced by the signal and ordinary electrons are practically identical 
 for the lateral shower development, the use of selection cuts on longitudinal shower development in the ECAL  results in  significant corrections for the signal efficiency. 
 \begin{table}[tbh!] 
\begin{center}
\caption{Expected signal efficiencies vs selection cuts calculated for the $\ainv$ decay of 
$A'$  with the mass $M_{A'} = 100$ MeV (see text for details).}
\label{tab:table1}
\begin{tabular}{lr}
\hline
\hline
Selection cut & Expected efficiency\\
\hline
Preshower $E_{PS}/E_0\gtrsim$  0.03& $ \gtrsim  0.95$\\
$E_{miss}/E_0\gtrsim $  0.5& $ \gtrsim  0.93$\\
ECAL X,Y matching& $ \gtrsim  0.90$\\
$e/\pi$ rejection, ECAL shower shape& $ \gtrsim  0.90$\\
VETO energy $E_{V}\lesssim 1$ MeV & $ \gtrsim  0.95$\\
HCAL energy   $ E_{HCAL}\lesssim  0.1$ GeV & $ \gtrsim  0.95$\\
\hline 
Total efficiency, $\epsilon_{eff}$  &         $ \gtrsim 0.65$\\
\hline
\hline
\end{tabular}
\end{center}
\end{table}

\subsection{The Veto and HCAL response to signal events}
One of the main variables defining the sensitivity of the experiment is 
the effective  width of the signal event distribution, shown in Fig.~\ref{sig-bckg}, 
along the $E_{HCAL}$-axis. The spread of the energy deposition of signal events 
in the HCAL is defined by the energy leak from the shower tail due to fluctuations  the longitudinal shower development and also by the admixture of the pule-up events.
The hadronic calorimeter is a set of four modules. Each module is a matrix of 3x3 cells. 
Each cell is a sandwich of alternating layers of steel and scintillator plates with thicknesses of 25 mm and 4 mm,  
respectively, and with a lateral size of $194\times 192$ mm$^2$.    Each cell consists of 48 such layers and has 
a total thickness of $\simeq 7\lambda_{int}$. 
The amount of the leak energy from the ECAL to the HCAL depends on the primary beam energy.
The thickness of the ECAL was chosen using the full shower simulation to minimize the amount of energy that leak into the 
Veto and HCAL. The purpose was to reduce it down to the level $\lesssim 100$ MeV (the  PED width of the HCAL electronics). In Fig.~\ref{ecalleak} the spectrum of the leak energy is shown. 

\section{Background}
The background reactions resulting in  the signature of Eq.(\ref{signinv}) can be classified as being due to physical-  and  beam-related sources. The discussions of these backgrounds are given in Refs. \cite{Gninenko:2013rka,Andreas:2013lya}. In this Section we consider several additional background sources   not studied  in Refs. \cite{Gninenko:2013rka,Andreas:2013lya} and show that their level is  below the expected 
sensitivity of the experiment.

\begin{itemize}
\item One possible source of background is caused by  the large transverse fluctuations of hadronic showers from the reaction  
\begin{equation}
e + Z \to e + Z + \geq 2~neutrals 
\label{eneutral}
\end{equation}
induced by electrons in the ECAL. 
In such events all secondary long-lived neutral particles (such as neutrons and/or $K^0_L$'s) could be  produced in the target at a large angle,   punch through the HCAL without depositing energy and escape the detector through the lateral surface resulting in the fake signal event. Note that background from events  with a leading neutral(s) is strongly suppressed by the HCAL thickness of $\simeq 30 \lambda_{int}$ in the 
forward direction.  
 
 The probability $P$ for  the reaction \eqref{eneutral} to occur  can be estimated as 
\begin{equation}
P \simeq  P_{n} \cdot P_{la} \cdot P_{leak}
\label{neutr}
\end{equation}
where $P_{n}, ~P_{la},~ P_{leak}$ are, respectively, the  fraction of the reaction \eqref{eneutral} per incoming beam electron, probability for production of energetic particles at large angle, the probability for these particles   to escape the HCAL without interactions. From the NA64 test run the fraction of events with a pure neutral hadronic  final state  in the reaction of 100 GeV electron scattering in the ECAL target  is found to be  $P_{n}\lesssim 10^{-6}$ per beam electron \cite{na64}. The $P_{leak}$ value   can be estimated as a probability for two neutrals with the total energy $\geq 50$ GeV-the threshold for $E_{miss}$ in the experiment- to escape HCAL by crossing  at least $\simeq 4\lambda_{int}$ each, under assumption that  both are produced at an angle of $\Theta_n \simeq 30^o$. This gives  $P_{leak}\lesssim 3\cdot 10^{-4}$.   

Because the cross section of the reaction \eqref{eneutral} is difficult to simulate, in order to estimate this  background we use the results of the NOMAD experiment which studied  large transverse fluctuations of  hadronic showers induced by  pions \cite{nomad1, nomad2}. In these  measurements the probability
 $P(f,R,E_\pi)$  to observe in an ECAL matrix a cluster  with the energy greater than a given fraction $f$ of the incoming pion with $E_\pi=15$  GeV, and at a distance $R$ from the beam axis has been measured. For example, the probability 
to find a separated cluster with the energy $ >0.1 E_\pi $ at a distance 30 cm (or $\Theta \gtrsim 30^o $) from the beam axis was found to be $P(f,R) \simeq 10^{-5}$ per incoming pion. 
The measurements also show that the probability $P(f,R)$ drops very quickly with increasing of the beam 
energy, $R$ ($\Theta_n$), or $f$.  E.g. for the same $f$ and $R$, the above $P$value is higher by a
 factor $\simeq 20$ for 6 GeV pions.   
 Neglecting this and also the difference in development of  hadronic showers  induced by pions and electrons,  
  we may consider the value  $P \simeq 10^{-5}$ as an upper limit on the  probability for the production of large angle   neutrals with energy $E_{n}> 0.1 \cdot E_0\simeq 5$ GeV at the beam energy $E_0\geq 50$ GeV.  Taking  this into account results  in a conservative estimate  for this background to be at the level 
 $\lesssim 10^{-14}$ per incoming electron. Note, that  the requirement to have two large angle neutrals carrying the total hadronic energy $\gtrsim 50$ GeV in the reaction \eqref{eneutral}, not $\geq 5$ GeV as discussed previously,  would significantly  suppress  background further. One may also consider more natural production angles smaller that $\simeq 30^o$. But in this case, the neutrals should pass without 
interaction longer distance $L$ in the HCAL and the probability $P_{leak}$ decreases quickly as 
$exp(-L/\lambda_{int})$. For example, if neutrals escape the first HCAL module just at its far end, the 
$P_{leak}\simeq 3\cdot10^{-7}$. Combining this with the probability $P_n$ results in $P$-value 
from Eq.(\ref{neutr}) already very  small, $P\lesssim 3\cdot 10^{-13}$.   
 
   Finally, we note that the presented estimate gives an illustrative order of magnitude for the background level from the large transverse fluctuations of hadronic showers produced in the reaction  \eqref{eneutral} and may be further improved either by more detailed simulations of the experimental setup, or by direct measurements similar to the NOMAD ones.


\item Another background can be due the electroproduction of di-muon pairs:  
\begin{equation}
e + Z \to e + \g + Z; ~ \g \to \mu^+ \mu^-,
\label{emuon}
\end{equation} 
 when the incident electron  produces in the ECAL target  a high-energy bremsstrahlung photon, which 
 subsequently  converts  into a $\mu^+ \mu^-$ pair in the field of the Pb nucleus. This process could mimic the signal either i) due to muons decay  in flight inside  the ECAL target  into $e\nu \nu$ state, or ii) if the muons escape  detection  in the V and HCAL modules due to  fluctuations of the energy   (number of photoelectrons) deposited  in these detectors.  For the case i)  the relatively long muon lifetime results  in a small probability to decay
 inside the ECAL. Assuming decay length of  $\simeq 20$ cm, a high suppression factor $\simeq 10^{-12}$ for this background source is calculated. Taking into account the additional suppression factor of 
 $\simeq 10^{-5}$ due to the cross-section  of the reaction \eqref{emuon} makes this background negligible. 
 For the case ii) the background is suppressed by the high-efficiency veto system V+HCAL. 
 The V is a  $\sim 1$  cm thick  high-sensitivity scintillator  arrays with a light yield  of  $\gtrsim 10^2$ photoelectrons per 1 MeV of deposited energy. The simulated distribution of energy deposited by muons in the V counter is shown in Fig. \ref{ecalleak}. It is also assumed that the veto inefficiency  for a single muon  detection  is, conservatively, $\lesssim 10^{-3}$ for the threshold $\simeq 0.5$ MeV ($\simeq 25$ photoelectrons). 
The number of photoelectrons produced by a MIP crossing the single  module  is in the range $\simeq$ 150-200 photoelectrons.  All these factors  lead to  the expectation for this  background to be  at the level at least $\lesssim 10^{-13}$ assuming 20-30 photoelectron threshold in the HCAL for two-MIP events.

\item The statistical limit on the sensitivity of the NA64 experiment is set  by the number of 
accumulated events which depend on the beam intensity. The intensity is limited by the ECAL signal duration ($\tau_{ECAL} \simeq 100$ ns) resulting in a maximally allowed electron counting rate  of $ \lesssim 1/ \tau_{ECAL} \simeq 10^{6}~ e^- /$s in order to avoid significant loss of the signal efficiency 
due to the pileup effect. To evade this limitation, one could implement a $e^-$-pileup removal algorithm to allow for high-efficiency  reconstruction of the signal shape and energy in high electron pileup environments,  and run the experiment at the electron beam rate $\simeq 1/\tau_{ECAL}\simeq$ a few $10^6~e^-/s$.

In addition, a random superposition of  uncorrelated low-energy,  50 - 70 GeV,  electron and 100 GeV 
pion (or muon) events  occurring during the detector gate-time could results in the following  fake signal. 
The low energy electron emits an amount of synchrotron radiation  energy which could  still be  above the detection threshold and then is deflected by the magnet so it does not hit the ECAL, see  Fig.~\ref{setup}. 
While the  accompanying  $\pi$ (or $\mu$) decays in-flight in front of the ECAL into the $e\nu$ ($e \nu \nu$) state with the decay electron energy less then the beam energy, thus resulting in the signal signature of Eq.(\ref{signinv}).  

 This background sources is related to the low-energy tail in the energy distribution of beam electrons. This tail is caused by the beam electron interactions with a passive material, such e.g. as entrance windows of the vacuum lines, residual gas, etc. in the upstream part of the beam line. Another source of low energy electrons is due to the $\pi$ or $\mu$ decays in flight. Taking into account that the fraction of such electrons  with energy   $50-70$ GeV in the 100 GeV electron  beam  could be as large as  $10^{-2}$, the time resolution of the $e^-$ and $\pi,\mu$ events is of the order of ns, the fraction of $\pi$ ($\mu$) in the beam is  $\lesssim 10^{-3}$ ($\lesssim 10^{-3}$), and the probability of the $\pi \to e \nu$ decay is $\simeq 2\cdot 10^{-7}$ ($\lesssim 10^{-6}$ for $\mu \to e \nu \nu$)  results in the level of this background to be  less than $10^{-15}$ ($10^{-14}$)  for $\pi$'s  ($\mu$'s)  per  electron for  the beam intensity    $\simeq 10^{6}~ e^- /$s.
\end{itemize}

\begin{figure}[tbh]
\begin{center}
\vskip-3.cm
\includegraphics[width=0.5\textwidth]{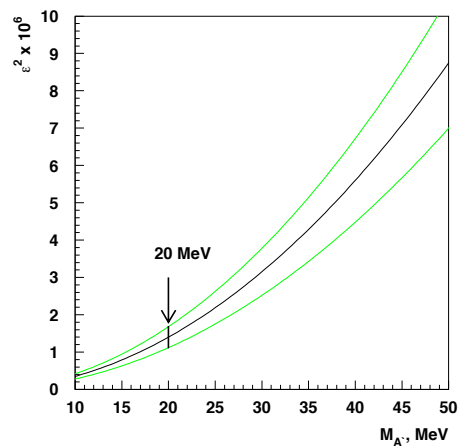}
\vskip-3.cm
\caption {{\small The dependence of the  mixing $\epsilon^2$ as a function of $\ma$ obtained  for 
$ \simeq 70$ observed events accumulated  with $\simeq 2\cdot 10^{10}$ eot. The green curves  represent the 90\% C.L..
Assuming that the observed events originated from decays of  $A'$ with mass  $M_{A'} = 20$ MeV would result in determination of  mixing strength interval around $\epsilon \simeq 10^{-3}$ value  indicated by the arrow. }}
 \label{epsma}
\end{center}
\end{figure} 
\begin{figure}[tbh]
\vskip-2.cm
\includegraphics[width=0.6\textwidth]{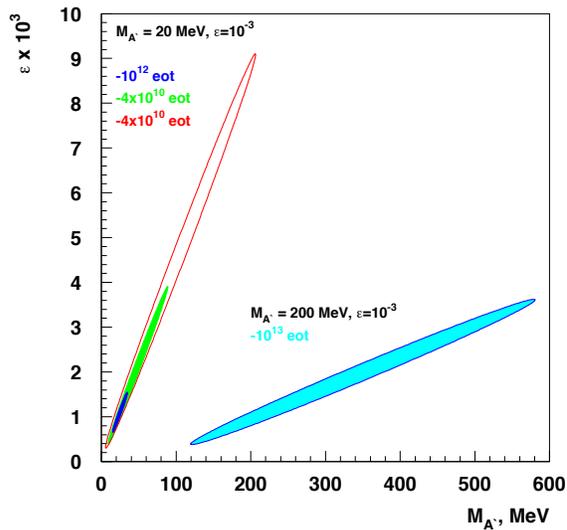}
\vskip-3.cm
\caption {\small
The fitted $\Delta \chi^2 =1 $ contours  in the $\epsilon$ vs $m_{A'}$ plane for invisibly decaying $A'$  obtained from the fit  of   $E_{miss}$ spectra calculated for the  $A'$ masses , $\ma=20$ MeV and  mixing $\epsilon = 10^{-3}$    and 
the missing energy $E_{miss}> 0.2 E_0$  for $4\cdot 10^{10}$ (green area) and $10^{12}$ (blue area ) eot, respectively. The red contour is calculated for the 
$E_{miss}> 0.5 E_0$ and  $4\cdot 10^{10}$  eot. For comparison, the fitted $\Delta \chi^2 =1 $ contour for $\ma=200$ MeV, 
$\epsilon \simeq 10^{-3}$, $E_{miss} >0.5 E_0$, and $n_{eot} \simeq 10^{13}$ is also shown.  
}
 \label{ellipse}
\end{figure}

\section{Expected results}

In this section we consider two possible outcomes of the experiment: A) observation of 
an excess of signal events associated with the reaction \eqref{reaction}, B) no excess of signal events is observed.

\subsection{Extraction of the  parameters $M_{A'}$ and $\epsilon$ using the missing energy spectrum.}
For the case of signal observation we performed a pseudo-experiment aiming at the study of the possibility  of extraction of the  parameters $M_{A'}$ and $\epsilon$.  
As an example, we consider values for the   $A'$ mass $M_{A'} = 20 $ and 200 MeV 
MeV  and mixing strength $\epsilon \simeq 10^{-3}$. Two possibilities were considered.
 
For the case of the $\lesssim 100$ signal events  observation  it would be 
possible to determine a band of allowed $\epsilon$ values in the two-dimensional plot ($\epsilon,~M_{A'}$).    
This could be done as follows. The observed number of signal events 
$n$ passing the selection cuts is distributed according to Poisson statistics 
\begin{equation}
P(n,\na)= \frac{\na^{n}}{n !} e^{-\na}
\end{equation}  
where $\na$ is the average number of signal events from the target.
The $\na$ depends in particular on $\epsilon,~M_{A'},~šE_e$, $n_{eot}$ - the total number of electrons on target, and is given by  
\begin{equation}
\na =n_{eot} \cdot \frac{\rho N_{A}}{A_{{\rm Pb}}} \cdot \epsilon_{eff}(\ma) \cdot 
{\int\limits_{E_0/2}^{E_0}}  \frac{d n}{d E_{A'}} \, d E_{A'} 
\label{AprYields}
\end{equation} 
where  $ \rho$ is density of Pb target, $N_A$ is the 
Avogadro's number, $A_{{\rm Pb}}$ is the  
Pb atomic mass, and $\epsilon_{eff}$ is the overall signal selection efficiency, see   
Table 1. The integration in Eq.~(\ref{AprYields}) is 
performed over the missing energy spectrum   in the  ECAL target, see Fig.\ref{AspectraForShower}. 
The equation Eq.(\ref{AprYields}) can by approximated by the form
\begin{equation}
\na = \frac{k\cdot n_{eot}}{10^{12}}\bigl(\frac{\epsilon}{10^{-5}}\bigr)^2 \bigl(\frac{10~\rm{MeV} }{M_{A'}}\bigr)^2   
\label{lambd}
\end{equation}  
where parameter $k$ weakly depends on $\ma$. For example,  for masses $M_{A'} = 20$ and 200 MeV, the $k$ values are 1.34 and 1.12,  resulting in the  yield $\na$(20 MeV)$=3.4\cdot10^3$ and $\na$(200 MeV)= 30 events, respectively, for $\epsilon=10^{-3}$,  $n_{eot}=10^{12}$ and $E_{miss}> 0.2 E_0$. 

\begin{figure}[tbh]
\begin{center}
\includegraphics[width=0.5\textwidth]{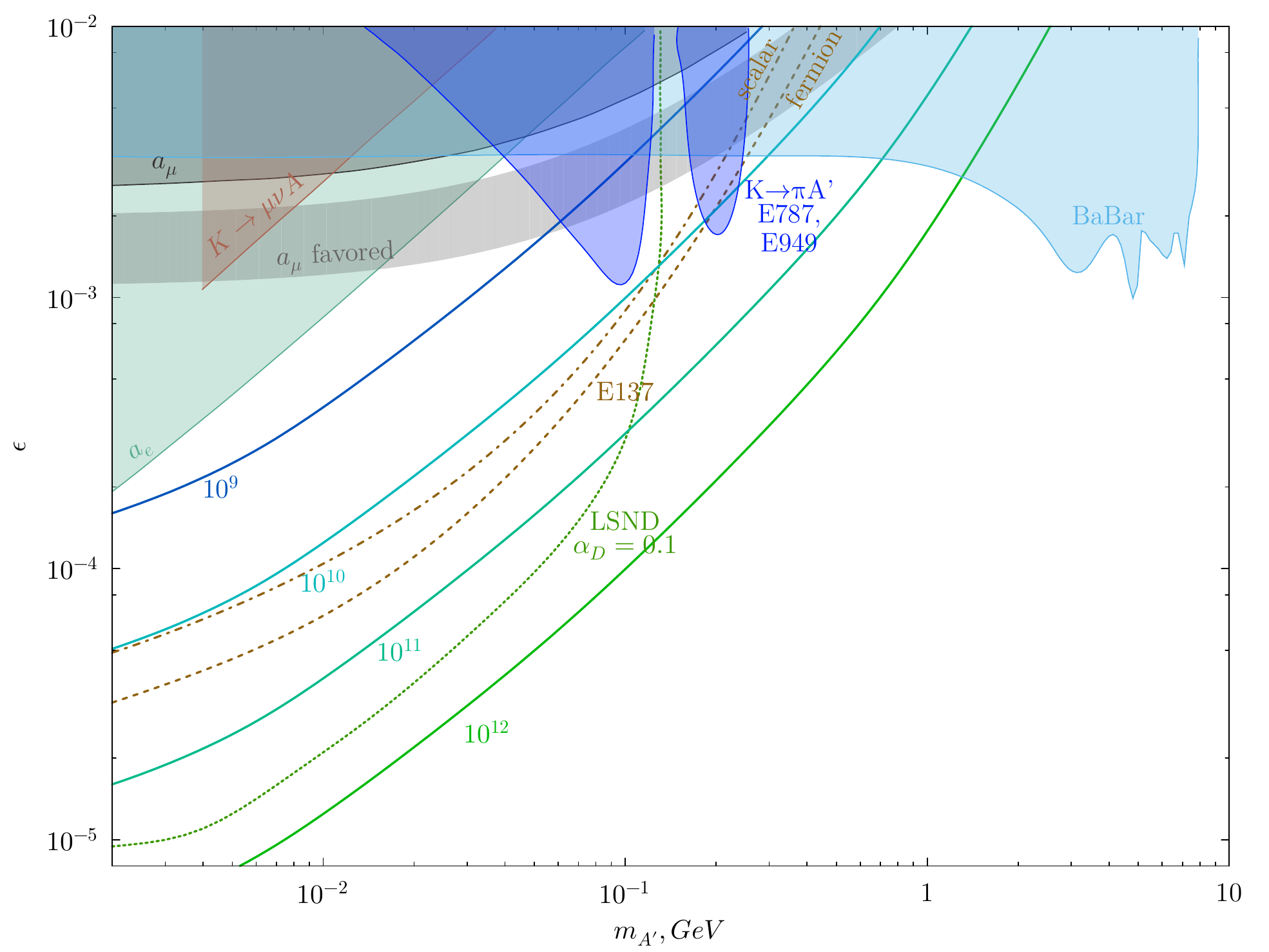}
\caption {
Expected exclusion region in the ($M_{A'}, \epsilon$) plane  
from the results of the proposed experiment for $10^{9}$, $10^{10}$, $10^{11}$ and 
 10$^{12}$ incident electrons at the 
 energy $E_{0}=100$  GeV. The curve are calculated under assumption that no background events 
 are observed for the given number of accumulated eot. 
  Direct constraints from the BaBar \cite{Izaguirre:2014bca, Aubert:2008as},   and E787+ E949  experiments \cite{Davoudiasl:2014kua, Artamonov:2009sz}, as well as muon (g-2) favored area 
 are also shown. The figures are based on Ref.\cite{Lee:2014tba}. Indirect constraints (95\% C.L.)  for
  dark photons $A'$ decaying invisibly to the pair of light DM $\chi$, extracted from the SLAC E137 \cite{Batell:2009di} for  a Dirac fermion or complex scalar ( broken brown) DM and from the LSND experiments \cite{deNiverville:2011it} (green dotted) under assumption  $\alpha_D = 0.1$ are also shown.
 For more limits obtained from indirect searches and 
 planned measurements see e.g. Refs. \cite{Essig:2013lka, raggi}.
  \label{exclinv}}
\end{center}
\end{figure}    
If  $\na \gg 1$ the Poisson distribution is approximated by the Normal 
distribution. Hence, for given $\epsilon,~šM_{A'}$ values, the number of signal events 
at "one-sigma" confidence level is given by
\begin{equation}
\na -\sqrt{\na} \leq n \leq \na +\sqrt{\na}
\label{lambda}
\end{equation}
Using  the expression \eqref{lambd} for  
the parameter $\na $ and  inequality \eqref{lambda},  
one  can estimate  from the data  the  ratio $\frac{\epsilon^2}{\ma^2}$.
An example of such estimate for  $\epsilon \simeq 10^{-3}$, $\ma=20$ MeV,  and $n_{eot} \simeq 2\cdot 10^{10}$ is shown in  Fig.~\ref{epsma}. In this case one can 
expect to observe $ \simeq 70 \pm 8$ signal events. 
 \begin{figure}[tbh]
\begin{center}
\includegraphics[width=0.5\textwidth]{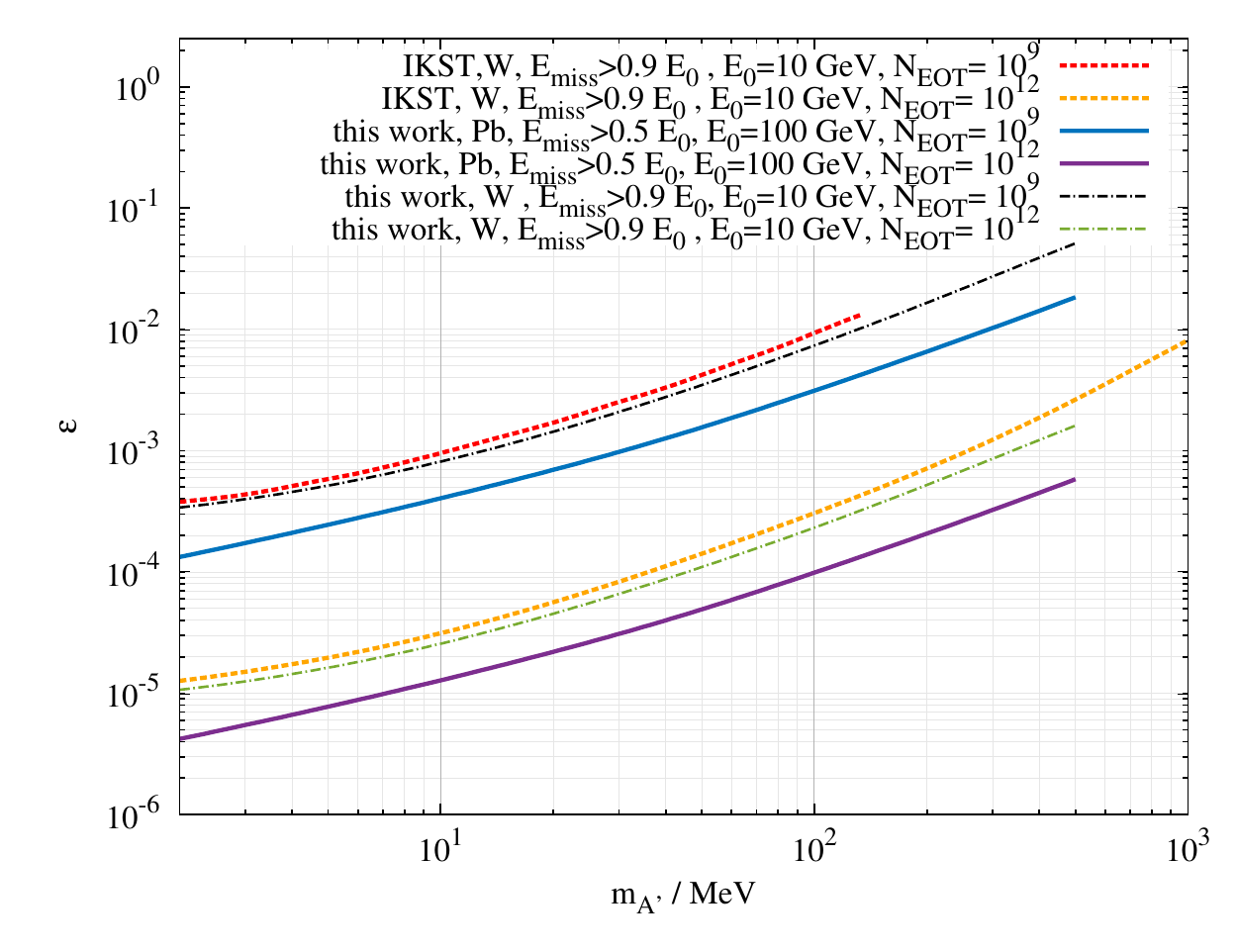}
\caption {{\small
Comparison of the upper limits in the $\epsilon$ vs $m_{A'}$ plane for invisibly decaying $A'$   calculated for the  W-ECAL target \cite{Andreas:2013lya}, $E_0 = 10$ GeV  ,  and 
the missing energy $E_{miss}> 0.9 E_0$
by  Izaguirre et al. \cite{Izaguirre:2014bca} (IKST)  and in this work  for $10^9$ eot  (red dashed and black  dash-dotted),   and $10^{12}$ eot (orange dashed and green dash-dotted), respectively. For comparison limits calculated for the shashlik ECAL target,  $E_0=100$ GeV ,  and the ECAL missing energy  $E_{miss} >0.5 E_0$  
for $10^9$ eot (blue solid)   and $10^{12}$ eot 
(purple solid), respectively, are also shown.  }
 \label{AprimeComprisonOfSensitivity}}
\end{center}
\end{figure}  

 For the  case of $n_{eot} \simeq 10^{12}$, $\ma \lesssim 50$ MeV  and larger number of the signal events  observation,  $n_{A'} \gtrsim 10^3$,   it should be possible to perform the more precise  measurement of the correlated parameters $\epsilon$ and $\ma$. 
This measurement is based on Eq.(\ref{TotalCrossSectionApr}) and the dependence of the shape
 of the missing energy spectrum on $\ma$, which is,  as shown in  Fig.~\ref{AspectraForShower}, 
 most significant in the region $E_{miss} \simeq 0.2$. 
 In this study, we assume that the $E_{miss}$ shape in this region is determined only by statistical errors. Two intervals of missing energy $0.5 E_0 < E_{miss} < E_0$ and $0.2 E_0 < E_{miss} < E_0$ were considered for comparison. 

Then, the following steps are made.
 On a grid of different $\epsilon$ and $\ma$ parameters for  each point we performed comparison of  the $E_{miss}$  distribution from  "observed" number of events with the simulated spectra. 
 The Kolmogorov and $\chi^2$  tests,   used for the shape compatibility check, give rather similar results. 
  The allowed regions with probabilities (p-values) expressed in terms of the corresponding numbers of standard deviations were finally obtained.  In Figure~\ref{ellipse} examples of  the one
standard deviation "ellipse" contours for the 
 best fit parameters for the different thresholds on $E_{miss}$ and numbers of 
accumulated eot is shown.  The best fit parameters are found to be 
 $M_{A'}=21.6$ MeV and $\epsilon=1.1\times 10^{-3}$ for the $n_{eot}\simeq 10^{12}$ collected electrons, which corresponds to  a few months of running. Note, that for higher masses $\ma \gtrsim 100$ MeV, the precision in determining of the   parameters $M_{A'}$ and $\epsilon$ for the given  value of $n_{eot}$ drops quickly with increasing of the mass $\ma$.   
 

 \begin{table*}[tbh!]
\begin{center}
  \begin{tabular}{|c|c|c|c|c|c|c|}
    \hline
    \multirow{2}{*}{ \!\!\!\!\!\! ~~$m_{A'}$,  \!\!\! MeV~~  \!\!\!\!} &
      \multicolumn{2}{c}{{\small $ {\bf (A)} $}} &
      \multicolumn{2}{|c|}{{\small ${\bf (B)}$}} &
      \multicolumn{2}{c|}{\small{${\bf (C)}$}}
    \\    \cline{2-7}
    &\!\!\!\! ~~$N_{eot}=10^9$~~ \!\!\!\! & \!\!\!\! ~~$N_{eot}=10^{12}$~~ \!\!\!\! &  \!\!\!\! ~~$N_{eot}=10^9$~~  \!\!\!\! &
     \!\!\!\! ~~$N_{eot}=10^{12}$~~  \!\!\!\! &  \!\!\!\! ~~$N_{eot}=10^9$~~  \!\!\!\! &  \!\!\!\! ~~$N_{eot}=10^{12}$~~  \!\!\!\! \\
    \hline
    2 & $1.33 \cdot 10^{-4}$  & $4.20 \cdot 10^{-6}$& $3.40 \cdot 10^{-4}$  & $1.07 \cdot 10^{-5}$   & $3.61 \cdot 10^{-4}$  & $1.20 \cdot 10^{-5}$  \\
    \hline
    10 & $3.91 \cdot 10^{-4}$  & $1.23 \cdot 10^{-5}$  & $8.14 \cdot 10^{-4}$  & $2.57 \cdot 10^{-5}$  & $8.98 \cdot 10^{-4}$  & $2.73 \cdot 10^{-5}$  \\
    \hline
    50  & $1.44 \cdot 10^{-3}$  & $4.57 \cdot 10^{-5}$  & $3.48 \cdot 10^{-3}$  & $1.10 \cdot 10^{-4}$  & $4.26 \cdot 10^{-3}$  & $1.29 \cdot 10^{-4}$  \\
    \hline
     500  & $1.84 \cdot 10^{-2}$  & $5.83 \cdot 10^{-4}$& $5.12 \cdot 10^{-2}$  & $1.61 \cdot 10^{-3}$ & $- $  & $2.77 \cdot 10^{-3}$  \\
    \hline
  \end{tabular}
  \caption{\label{EpsilonNumValues}
  Upper bounds on mixing $\epsilon$ at 90
      \% CL for the following cases:  ${\bf (A)} $: this work,
       Pb-Sc dump,  $E_{miss} > 0.5 E_0$, $E_0=100$ GeV; $ {\bf (B)} $: 
      this work, W-Sc dump, $E_{miss} > 0.9 E_0$, ${\bf E_0= 10}$ GeV; $ {\bf (C)} $:  IKST, W-dump, $E_{miss} > 0.9 E_0$, ${\bf E_0= 10}$ GeV. 
  }
  \end{center}
\end{table*}
 
\subsection{Expected sensitivity }

In this section we consider expected  bounds on dark  
photon  parameter space based on the {\tt GEANT4} MC simulation of the $A'$  yields in the 
NA64 experiment.
 We define the acceptance of the detector 
$\eta_{acc}$ as the ratio of signal events with the missing energy 
$E_{miss}>0.5 E_0$ to the total 
number of events with a dark photon emitted in the  target. All bounds are calculated under assumption 
that no background events are observed for the given number of accumulated eot. 

Using Eq.(\ref{AprYields}) and the relation $n_{A'}^{90\%} > \na$, 
where $n_{A'}^{90\%}$ is the 90$\%~CL$ upper limit for the  number of 
signal events  without background, 
$n_{A'}^{90\%}  = 2.3$, one can  determine 
the expected $90\%~ CL$ bounds on $(M_{A'}, 
\epsilon)$ parameter 
space, which are shown in 
Fig.~\ref{exclinv}. The bounds are obtained for the total number of electrons on target $n_{eot} = 
 10^9, 10^{10},  10^{11}$, and $10^{12}$ and   the electron beam  energy 
 $E_0 = 100$ GeV. We assume  that the $A'$s decays dominantly to the invisible final state.
In Fig.~\ref{AprimeComprisonOfSensitivity} and Table II  we show detailed  
comparison of the  expected  sensitivity for the $A'$ invisible 
decay search  in our experiment calculated in this work with the one 
evaluated by  Izaguirre et al. in  Ref.\cite{Izaguirre:2014bca} for the 
case of the  W-Sc ECAL  and  $10^{9}$ eot. The comparison  is made for the case of the same type of the ECAL
(the W-Sc sandwich calorimeter \cite{Gninenko:2013rka, Andreas:2013lya}) , the beam energy $E_0 = 10$ GeV,
 the missing energy range  $E_{miss} >  0.9 E_0$ and for $10^{9}$ and  $10^{12}$ eot. Our results for the 
case of the Pb-Sc (shashlik) ECAL, the beam energy $E_0 = 100$ GeV, the missing energy $E_{miss} > 0.5 E_0$ and $n_{eot}=10^{9},10^{12}$ eot
 are also shown for comparison.  
For the former case, the expected bounds for 
 tungsten ECAL are in agreement with IKST  limits within 10 \%. 
 In Tab.~\ref{EpsilonNumValues}  we show the expected limits on mixing $\epsilon$
 at 90 \% CL  for the relevant  benchmark masses  $m_{A'}$
 and ECAL energy thresholds. For the second case, one can see that the sensitivity is two times better than for the former one.
 This is mainly due to the extension of the allowed missing energy region from $0.5E_0<E_{miss}<E_0$
 to $0.9 E_0<E_{miss}<E_0$ for signal events.

\section{Summary}

In this Section, we briefly outline the main improvements achieved in this article with respect to our previous work, as well as the recent work carried out by
another group. We have studied  the missing energy signature of the production of sub-GeV dark photons in the process of high-energy electron scattering off nuclei 
   in the experiment NA64 aiming at the search for $\ainv$  decays  at the CERN SPS.    We have shown  the distinctive  distributions of these events that
serve to distinguish the $\ainv$ signal from background.   
The results of the detailed simulations of the detector response  and efficiencies  to the signal events are presented. The comparison of the lateral shower profiles  for the electron and signal events  in the ECAL show that they are identical with high accuracy. 
No significant difference is found. While the longitudinal development of the electron and signal induced showers in the ECAL is quite  different. Thus a special 
attention is required to the selection of a threshold for the energy deposited in the 
preshower to keep the signal efficiency as high as possible.   

Using these results we evaluate the expected sensitivity of the experiment and show that it potentially allows  to  probe the still unexplored area of the mixing
strength  $10^{-6}\lesssim \epsilon \lesssim 10^{-2}$ and masses $M_{A'} \lesssim 1$ GeV.
The results obtained are found to be in agreement with the results of Ref.\cite{Izaguirre:2014bca} obtained for the same experimental setup
and selection criteria. For a realistic study of the expected 
sensitivity of the experiment we have improved on two points: 
we employed the $A'$ production into the GEANT4 simulation package, and 
performed the full simulation of the detector response to the 
$\ainv$ signal events. 
 We re-checked the results of Ref.\cite{Izaguirre:2014bca} where the $A'$
yield was carefully derived, and improved it further by taking into account 
the simulation of the realistic detector configuration, the detector response 
and the corresponding efficiencies. 
We believe that the error of the estimates of the experiment sensitivity 
obtained in  in those two works is unlikely exceed 10\%, which could be 
 attributed to  the uncertainty of the $A'$ yield. Taking as a banchmark 
the $M_{A'}=20$ and 200 MeV and $\epsilon=10^{-3}$ values 
we have determined   these parameters by 
 fitting Monte Carlo simulated $E_{miss}$ distributions.
 The best fit parameters are found to be 
 $M_{A'}=21.6$ MeV and $\epsilon=1.1\times 10^{-3}$ for the $n_e \simeq 2\cdot 10^{12}$
accumulated eot. We also determined the $\Delta \chi^2 = 1$ contours in the
($M_{A'} ; \epsilon$) parameter space and 
 demonstrated that in the case 
 of signal observation  estimated sensitivity of the search allows to determine its  parameters  with precision which strongly depends on the number of accumulated eot.

{\large \bf Acknowledgments}

We  would like to  thank members of the NA64 Collaboration for the numerous
 discussions, and,  
in particular  R. Dusaev and B. Vasillishin for their help.  
We are grateful to  the authors of Ref.\cite{Izaguirre:2014bca} for letting us  now their results prior  the publication, and in particular to P. Schuster  for useful comments.  The work of D.K. on simulations of signal events has been supported by the RSCF grant 14-12-01430.

\end{document}